\documentclass[a4paper,aps,prd,floatfix,reprint,twoside,twocolumn,preprintnumbers,showkeys,showpacs,final]{revtex4-2}
\usepackage[utf8]{inputenc}
\usepackage{amsmath, amssymb}
\usepackage{siunitx}
\usepackage{graphicx}
\usepackage[caption=false]{subfig}
\usepackage[section]{placeins}
\usepackage{verbatim}

\usepackage{booktabs}
\usepackage{dcolumn}
\usepackage{enumitem}

\usepackage{hyperref} 

\usepackage{environ}
\usepackage{tikz}
\usetikzlibrary{arrows,decorations.markings}
\usetikzlibrary[arrows.meta,bending]
\usetikzlibrary{positioning}

\makeatletter \newsavebox{\measure@tikzpicture}
\NewEnviron{scaletikzpicturetowidth}[1]{%
  \def\tikz@width{#1}%
  \begin{lrbox}{\measure@tikzpicture}%
    \BODY
  \end{lrbox}%
  \pgfmathparse{#1/\wd\measure@tikzpicture}%
  \BODY } \makeatother

\DeclareMathOperator{\Tr}{Tr}

\newcommand{\mc}[1]{\multicolumn{1}{c}{$#1$}}
\newcommand{\cdash}{\multicolumn{1}{c}{$-$}}

\begin{document}
\preprint{ADP-22-11/T1182}

\title{Static quark potential from centre vortices in the presence of dynamical fermions}
\author{James C. Biddle} \author{Waseem Kamleh} \author{Derek B. Leinweber}
\affiliation{Centre for the Subatomic Structure of Matter, Department of Physics, The University of Adelaide, SA 5005, Australia}

\begin{abstract}
  For the first time, centre vortices are identified on SU(3) lattice ensembles that
  include dynamical fermions. Using a variational method, the static quark potential is
  calculated on untouched, vortex-removed, and vortex-only fields. Two dynamical ensembles and one pure gauge ensemble are studied, allowing for an exploration of the impact of
  dynamical fermions on the centre-vortex vacuum. Novel modifications to the standard Coulomb term are introduced to describe the long range behaviour of the vortex-removed potential. These modifications remove a source of systematic error in the fitted string tension on the original ensembles. Our pure Yang-Mills result is consistent with previous studies, where projected centre-vortex fields only reproduce approximately two thirds of the string tension. Remarkably, we find that the vortex-only fields on both dynamical lattices are able to fully reproduce the respective untouched string tensions.
\end{abstract}

\maketitle

\section{Introduction}\label{sec:Intro}

Over recent years, centre vortices have
been shown to play a pivotal role in the generation of dynamical chiral symmetry
breaking and quark confinement in the QCD vacuum~\cite{tHooft:1977nqb,tHooft:1979rtg,DelDebbio:1996lih,Faber:1997rp,DelDebbio:1998luz,Bertle:1999tw,Faber:1999gu,Engelhardt:1999fd,Engelhardt:1999xw,Engelhardt:2000wc,Bertle:2000qv,Langfeld:2001cz,Greensite:2003bk,Bruckmann:2003yd,Engelhardt:2003wm,Boyko:2006ic,Ilgenfritz:2007ua,Bornyakov:2007fz,OCais:2008kqh,Engelhardt:2010ft,Bowman:2010zr,OMalley:2011aa,Trewartha:2015ida,Trewartha:2015nna,Greensite:2016pfc,Trewartha:2017ive,Biddle:2018dtc,Spengler:2018dxt}. In pure-gauge QCD, it has been
shown that vortex removal results in a loss of dynamical mass
generation~\cite{Trewartha:2015nna,OMalley:2011aa,Trewartha:2017ive}, loss of
string tension~\cite{Langfeld:2003ev,Bowman:2010zr} and the suppression of the
infra-red Landau gauge gluon propagator~\cite{Biddle:2018dtc,Bowman:2010zr}.
However, quantitatively reproducing these properties from vortex-only fields has
proved elusive. In studies performed on pure Yang-Mills $SU(3)$ gauge fields,
it is well known that vortices alone can only account for $\sim 62\%$ of the full string
tension~\cite{Langfeld:2003ev,OCais:2008kqh,Trewartha:2015ida}. Similarly, the Landau gauge quark and gluon propagators calculated on vortex-only
fields do not agree with their original values except after
smoothing~\cite{Trewartha:2015nna,Biddle:2018dtc}.

A natural next step for the vortex model is to examine how the presence of
dynamical fermions impacts the structure of centre vortices. Any subsequent
shift in vortex structure can be measured by calculating observables arising
from vortex-only and vortex-removed ensembles. In this paper, we perform the
first such analysis and present a calculation of the static quark potential on
vortex-modified ensembles in the presence of dynamical fermions. After
identifying centre vortices on the lattice, it is possible to isolate the
contribution to the static quark potential from both the vortices alone and the
original gauge field after vortex removal. This calculation reveals a
significant shift in vortex structure induced by the presence of fermion loops
in the vacuum fields and further reinforces the central role vortices play in
producing the salient phenomena of QCD.

This paper is structured as follows. Section~\ref{sec:VID} outlines how centre
vortices are identified on the lattice. Section~\ref{sec:SQP} introduces the
calculation of the static quark potential through use of Wilson loops.
Section~\ref{sec:VarAnalysis} describes the variational method used to
calculate the static quark potential. Section~\ref{sec:Results} discusses the
results of this work, introducing novel modifications to the standard Coulomb term. Section~\ref{sec:Conclusion} summarises our
findings.

\section{Vortex identification}\label{sec:VID}

In the continuum, centre vortices are regions of an $SU(N)$ gauge field that
carry flux associated with the centre of the gauge group. These regions are
`thick', meaning that in four dimensions they appear as three-dimensional
volumes. On the lattice, we instead identify `thin' vortices that are correlated
with the location of the physical thick
vortices~\cite{Engelhardt:1999xw,Bertle:2000ap}. These thin vortices are
two-dimensional sheets in four dimensions, which, when projected to three
dimensions, appear as closed loops. Visualisations of these centre vortices on the
lattice have been presented in Ref.~\cite{Biddle:2019gke}.

To identity centre vortices on the lattice, we first transform each
gauge field configuration to maximal centre gauge (MCG). This is done by finding the gauge
transformation $\Omega(x)$ that serves to maximise the
functional~\cite{Langfeld:2003ev,Trewartha:2015ida}

\begin{equation}
  \label{eq:MCGGlobal}
  R = \frac{1}{V\, N_\text{dim}\, n_c^2} \sum_{x, \mu}\left| \Tr U^{\Omega}_{\mu}(x) \right|^2\, .
\end{equation}
This gauge transformation brings each link as close as possible to the centre of
the $SU(3)$ group. For $SU(3)$, the centre of the group contains the three elements
\begin{equation}
  \mathbb{Z}_3 =  \left\{ \exp\left(\frac{m\, 2\pi i}{3} \right)I, ~ m= 0, \pm 1 \right\}.
\end{equation}
After fixing to maximal centre gauge, the nearest centre element is defined by
finding the minimum difference in phase between $\Tr U_{\mu}(x)$ and one of the
elements of $\mathbb{Z}_3$. $U_{\mu}(x)$ can then be projected onto this nearest
centre element to obtain the vortex-only configurations, $Z_{\mu}(x)$. The
vortex-removed configurations are then defined as
$R_{\mu}(x) = Z_{\mu}^{\dagger}(x)\, U_{\mu}(x)$.

For this work we make use of three ensembles of 200 $32^3\times 64$ lattice
gauge fields. Two of these are $(2 + 1)$ flavour dynamical ensembles from the
PACS-CS collaboration~\cite{Aoki:2008sm}. We choose the heaviest and lightest
pion mass ensembles to provide the greatest differentiation as the physical
point is approached. The pure gauge ensemble was generated with the Iwasaki
action~\cite{Iwasaki:1983iya} at $\beta=2.58$ with the intent of having a
similar lattice spacing as the PACS-CS ensembles. This allows us to readily
compare the full QCD results with those obtained from the pure gauge ensemble.
%

\begin{figure}[!tbp]
  \centerline{\includegraphics[width=0.9\linewidth]{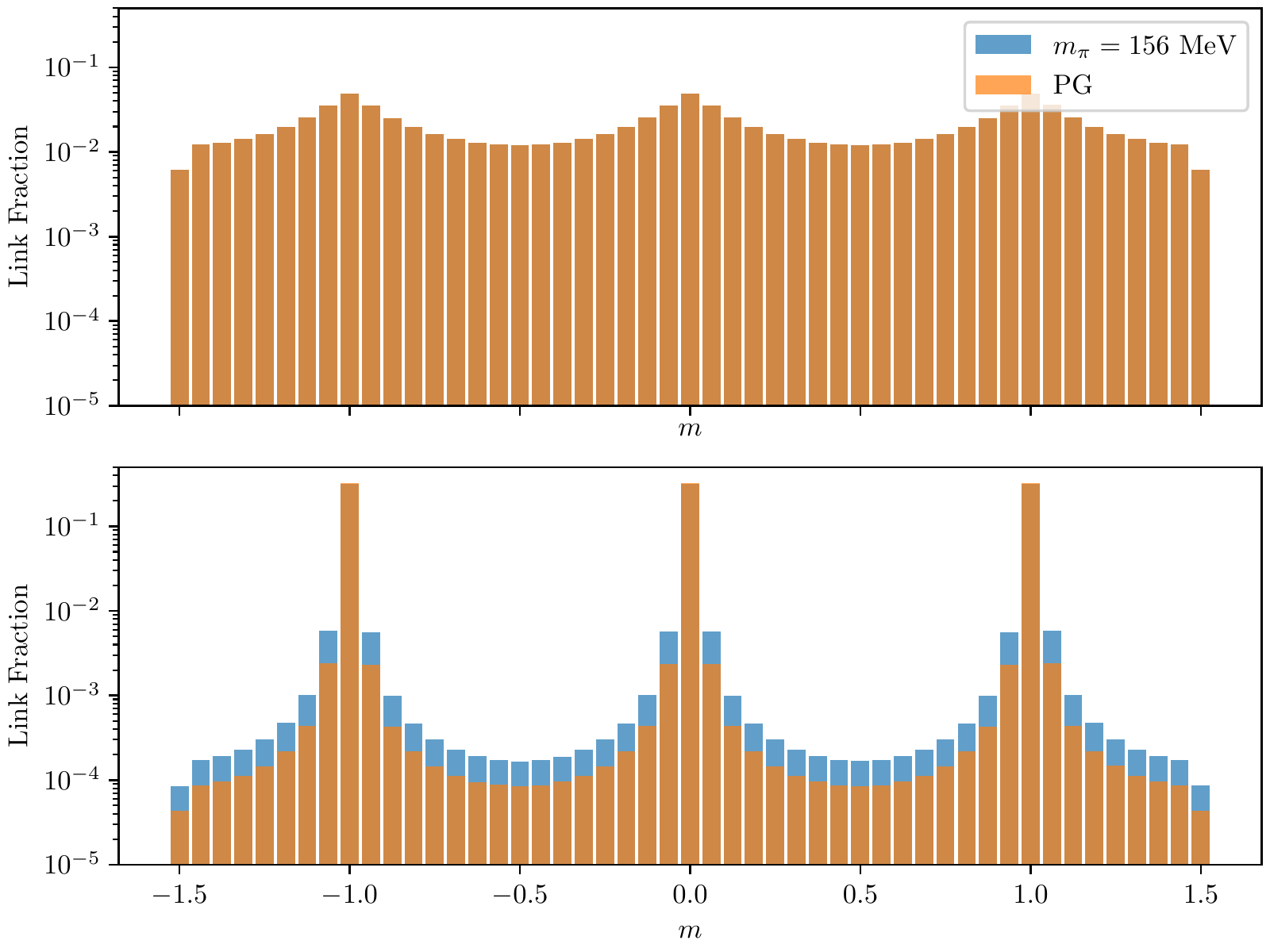}}
  \caption{\label{fig:MCGPhases} A histogram showing the average phase
    distribution $m$ of the pure gauge and lightest pion mass ensembles before
    ({\bf top}) and after ({\bf bottom}) maximal centre gauge fixing. Note the
    logarithmic scale. In the top plot, the agreement is so close that the
    dynamical ensemble results are hidden by the pure gauge.}
\end{figure}

For each of these lattices, the MCG procedure above creates a corresponding set of vortex-modified fields. Throughout the rest of this work we refer to the three field types derived from a lattice ensemble as the:
\begin{itemize}
\item Original, untouched (UT) fields, $U_{\mu}(x)$,
\item Vortex-only (VO) fields, $Z_{\mu}(x)$, and
\item Vortex-removed (VR) fields, $R_{\mu}(x)$.
\end{itemize}
The effectiveness of the MCG procedure can be seen in Fig.~\ref{fig:MCGPhases},
which shows a histogram of centre phases before and after MCG fixing on the pure
gauge and lightest pion mass dynamical ensembles. Interestingly, we find that
the pure gauge ensemble is more strongly peaked around the centre phases,
although the discrepancy is small, made visible by the logarithmic scale. A
summary of the ensemble parameters can be found in
Table~\ref{tab:LatticeParams}.
\begin{table}[tb]
  \caption{\label{tab:LatticeParams}A summary of the lattice ensembles used in
    this work~\cite{Aoki:2008sm}.}
  \begin{ruledtabular}
    \begin{tabular}{lcccc}
      Type & \mc{a\, (\si{fm})} & \mc{\beta} & \mc{\kappa_{\rm u,d}} & \mc{m_{\pi}\, (\si{MeV})} \\
      \midrule
      Pure gauge & 0.100 & 2.58 & \cdash & \cdash \\
      Dynamical & 0.102 & 1.90 & 0.13700 & 701\\
      Dynamical & 0.093 & 1.90 & 0.13781 & 156
    \end{tabular}
  \end{ruledtabular}
\end{table}
%

\subsection{Parallel MCG fixing}

Given the size of the lattices used in this work, it was necessary to implement
a parallel version of the MCG algorithm, which proceeds as follows.
To construct the maximal centre gauge transformation $\Omega(x)$, it is
sufficient to consider the nearest-neighbour contributions from $U_{\mu}(x)$ and
$U_{\mu}(x - \hat{\mu})$ $\forall\, \mu ~ \in \lbrace 1, 2, 3, 4\rbrace$. For each $x$,
one then seeks to maximise the local functional~\cite{Montero:1999by}
\begin{align}
  \label{eq:MaxFunc}
  R(x) = &\sum_{\mu}\left| \Tr \Omega(x)\,U_{\mu}(x)\right|^2 + \nonumber\\
  & \sum_{\mu}\left| \Tr  U_{\mu}(x-\hat{\mu})\,\Omega(x)^{\dagger}\right|^2
\end{align}
This is achieved by considering each of the three $SU(2)$ subgroups of $SU(3)$.
$\Omega(x)_{{\rm SU(2)}}$ is then expressed as a linear combination of the
$SU(2)$ generators $\vec{\sigma}$ such that
\begin{equation}
  \Omega_{\rm SU(2)}(x) = g_{4}I - i\vec{g}\cdot \vec{\sigma}.
\end{equation}
This reduces Eq.~\ref{eq:MaxFunc} to a quadratic in $(g_4,\, \vec{g})$ subject to a unitarity
constraint that can then be minimised via standard Lagrangian multiplier
techniques. Once each of the three $SU(2)$ subgroups is iterated over once and
$\Omega(x)$ has been constructed, it is then applied to the nearest-neighbour
gauge links. The process is repeated for all other values of $x$ and then
iterated until a plateau in $R$ (see Eq.~\eqref{eq:MCGGlobal}) is reached.

\begin{figure}
  \begingroup
  \tikzset{every picture/.style={scale=0.7}}%
      \begin{tikzpicture}[>={Latex[length=2mm]}, blackarr/.style={->,shorten >=
        5pt, shorten <= 5pt}, cyanarr/.style={->,cyan,shorten >= 2pt, shorten <=
        2pt}, bigcyanarr/.style={->,cyan,shorten >= 2pt, shorten <= 2pt},
      redarr/.style={->,red,shorten >= 2pt, shorten <= 2pt}]

      \coordinate (start) at (0,0);
      \coordinate (mids1) at (1,0);
      \coordinate [label={[red, label distance=5pt]90:$3$}] (x1) at (2,0);
      \coordinate (mid12) at (3,0);
      \coordinate [label={[cyan, label distance=5pt]90:$1$}](x2) at (4,0);
      \coordinate (mid2proc) at (5,0);
      \coordinate (proc) at (6,0);
      \coordinate [label={[align=center]above:Processor\\boundary}] (up) at (6,2);
      \coordinate (down) at (6,-2);
      \coordinate (midproc3) at (7,0);
      \coordinate [label={[red, label distance=5pt]90:$3$}] (x3) at (8,0);
      \coordinate (mid34) at (9,0);
      \coordinate [label={[cyan, label distance=5pt]90:$1$}] (x4) at (10,0);
      \coordinate (mid4e) at (11,0);
      \coordinate (end) at (12,0);

      \filldraw[red] (x1) circle (2pt); \filldraw[cyan] (x2) circle (2pt);
      \filldraw[red] (x3) circle (2pt); \filldraw[cyan] (x4) circle (2pt);

      \draw[cyanarr] (x2) to [bend right=80] node [midway,above]{} (mid12);
      \draw[cyanarr] (x2) to [bend left=80] node [midway,above]{} (mid2proc);

      \draw[cyanarr] (x4) to [bend right=80] node [midway,above]{} (mid34);
      \draw[cyanarr] (x4) to [bend left=80] node [midway,above]{} (mid4e);

      \draw[redarr] (x1) to [bend right=80] node [midway,above]{} (mids1);
      \draw[redarr] (x1) to [bend left=80] node [midway,above]{} (mid12);

      \draw[redarr] (x3) to [bend right=80] node [midway,above]{} (midproc3);
      \draw[redarr] (x3) to [bend left=80] node [midway,above]{} (mid34);

      \draw[cyanarr] (mid2proc) to [bend right=80] node
      [midway,above,label={[below]-45:$2$}]{} (midproc3); \draw[cyanarr] (mid4e)
      to [bend left=30] node [midway,above,label={[below]-45:$2$}]{} (mids1);

      \draw[redarr] (mids1) to [bend right=50] node [midway,above,label={[below]-45:$4$}]{} (mid4e);
      \draw[redarr] (midproc3) to [bend right=80] node [midway,above,label={[above]-45:$4$}]{}
      (mid2proc);

      \draw [blackarr,dashed,very thick](start) -- (x1) ;
      \draw [blackarr,very thick](x1) -- (x2) ;
      \draw [blackarr,very thick](x2) -- (proc) ;
      \draw [blackarr,dashed,very thick](proc) -- (x3) ;
      \draw [blackarr,very thick](x3) -- (x4) ;
      \draw [blackarr,very thick](x4) -- (end) ;

      \draw [dashed] (down) -- (up);

    \end{tikzpicture}
  \endgroup
  \caption{\label{fig:MCGAlg} MCG updating scheme for two processors. The update
    process is described in the text.}
\end{figure}
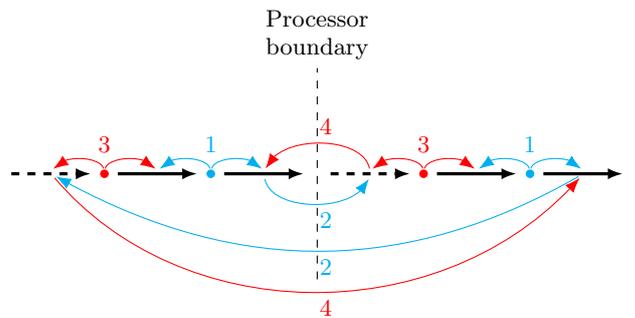

 As $\Omega(x)$ depends only on its
nearest-neighbours, we mask the algorithm to ensure that at any one time we
consider only even or odd values of $x$, where even or odd is defined by whether
$\sum_{\mu = 1}^{4}x_{\mu}$ is even or odd. We then distribute regular chunks of
the lattice across processors with one shadowed plane in the directions along
which the lattice has been subdivided. Once an even or odd sweep has been
completed, the updated links are copied to adjacent processors so that they are
available for the alternate sweep. A diagram illustrating this updating scheme
for two processors distributed along one dimension is shown in Fig.~\ref{fig:MCGAlg}.

The processor boundary is shown with the vertical dashed line. Gauge links are
shown with solid black arrows and shadowed gauge links are shown with black
dashed arrows. Shown is the update process starting with the even sites
(blue circles) followed by the odd sites (red circles):
\begin{enumerate}
\item The gauge links adjacent to the even sites are updated with the gauge
  transformation $\Omega(x)$.
\item The updated links along the boundary are
  copied to the relevant shadowed locations.
\item  The gauge links adjacent
  to the odd sites are updated.
\item The updated shadowed links are copied
  to the relevant locations.
\end{enumerate}%
This method of parallel implementation requires a slightly greater number of
overall sweeps than the serial implementation, as each update does not have the
fully propagated information that would be carried by a serial process starting
from one corner of the lattice. However, it has a number of advantages. Most
apparent is the real-time reduction in wall time, as the
parallel implementation scales very well thanks to minimal cross-processor
memory requirements. Additionally, there is no directionality in this
implementation as each site only sees its neighbours during each sweep. This
suppresses any inconsistency arising from choice of start point or order of
iteration. Given that each site is only affected by its nearest neighbours,
this implementation also has the desirable property of being agnostic to the
number of processors used in the calculation.


\section{Static Quark Potential}\label{sec:SQP}
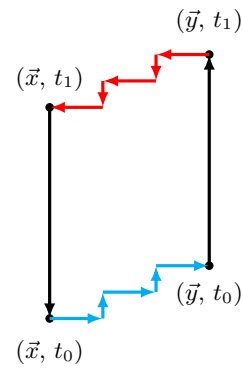
\begin{figure}[!t]
  \begingroup
  \tikzset{every picture/.style={scale=0.7}}
  \begin{tikzpicture}[>={Latex[length=2mm]}, blackarr/.style={->},
  cyanarr/.style={->,cyan},
  bigcyanarr/.style={->,cyan,shorten >= 2pt, shorten <= 2pt},
  redarr/.style={->,red}]

\coordinate [label={[label distance=5pt]-90:$(\vec{x}, \, t_0)$}] (bl) at (0,0);
\coordinate [label={[label distance=3pt]90:$(\vec{x}, \, t_1)$}] (tl) at (0,4);
\coordinate [label={[label distance=5pt]90:$(\vec{y}, \, t_1)$}] (tr) at (3,5);
\coordinate [label={[label distance=3pt]-90:$(\vec{y}, \, t_0)$}] (br) at (3,1);

\coordinate (th1) at (1,4);
\coordinate (tv1) at (1,4.5);
\coordinate (th2) at (2,4.5);
\coordinate (tv2) at (2,5);

\coordinate (bh1) at (1,0);
\coordinate (bv1) at (1,0.5);
\coordinate (bh2) at (2,0.5);
\coordinate (bv2) at (2,1);

\coordinate (bc) at (1.5,0.5);
\coordinate (tc) at (1.5,4.5);

\filldraw[black] (bl) circle (2pt);
\filldraw[black] (br) circle (2pt);
\filldraw[black] (tl) circle (2pt);
\filldraw[black] (tr) circle (2pt);

\draw [blackarr,very thick](tl) -- (bl) ;
\draw [blackarr,very thick](br) -- (tr) ;


\draw[redarr, very thick] (th1) to (tl);
\draw[redarr, very thick] (tv1) to (th1);
\draw[redarr, very thick] (th2) to (tv1);
\draw[redarr, very thick] (tv2) to (th2);
\draw[redarr, very thick] (tr) to (tv2);

\draw[cyanarr, very thick] (bv2) to (br);
\draw[cyanarr, very thick] (bh2) to (bv2);
\draw[cyanarr, very thick] (bv1) to (bh2);
\draw[cyanarr, very thick] (bh1) to (bv1);
\draw[cyanarr, very thick] (bl) to (bh1);


\end{tikzpicture}
  \endgroup
  \caption{\label{fig:WilsonLoop} Diagram of a Wilson loop. Shown are the forward (blue) and backward (red) spatial paths where different levels of smearing are used to create our variational matrix. Links in the positive temporal direction are oriented vertically upwards.}
\end{figure}

The static quark potential provides a measurement of the potential between two
massive, static quarks at a separation distance $r$. On the lattice, the static
quark potential can be obtained by considering the Wilson loop%
\begin{equation}
  W(r, \, t) = \Tr R(\vec{x},t_0)T(\vec{y},t_0)R^\dagger(\vec{x},t_1)T^\dagger(\vec{x},t_0)
\end{equation}
that has two spatial paths connecting points $\vec{x}$ and $\vec{y}$ satisfying
$| \vec{y} - \vec{x} | = r$ via the shortest set of links on the lattice.
The forward spatial path $R(\vec{x},t_0)$ is separated from the backward
spatial path $R^\dagger(\vec{x},t_1)$ by the temporal extent of the loop, $t_1-t_0=t.$ The loop is closed via the static quark propagators $T(\vec{y},t_0)$ and $T^\dagger(\vec{x},t_0),$ which correspond to the product of links in the positive and negative temporal directions respectively. A diagram of this Wilson loop construction is shown in Fig.~\ref{fig:WilsonLoop}.

When the spatial separation extends off-axis to encompass displacements in more than one spatial direction, a diagonal path is chosen to reduce rotational lattice artefacts. An integer step size vector $\vec{s}$ is initialised by taking the spatial separation $\vec{r}$ and dividing out the smallest element. If the two largest elements of $\vec{s}$ are both greater than 1, then these are divided by the smaller of the two so that the step size vector $\vec{s}$ has at most one element that is greater than 1. The spatial link path is constructed by cycling between the spatial directions $\hat{\jmath}$ with step size $s_j.$ When the total displacement $r_j$ in a direction $\hat{\jmath}$ has been reached we set the step size $s_j = 0.$  This is perhaps most easily understood with an example. For $\vec{r} = (6,3,2),$ then the initial step size vector $\vec{s} = (3,1,1).$ The path $\vec{r}$ is traversed by starting at $\vec{x}$ and cycling through the steps $\vec{s} = (3,1,1)$ twice, then updating $\vec{s} = (0,1,0)$ to the remaining displacement to reach the end point $\vec{y}.$  

The expectation value of the Wilson loop is connected to the static quark potential
$V^{\alpha}$ for state $\alpha$ via the expression
\begin{equation}
  \label{eq:SQP}
  \langle W(r, \, t) \rangle = \sum_\alpha \lambda^{\alpha}(r)\, \exp \left(-V^{\alpha}(r)\, t\right)\, .
\end{equation}
Here, $\alpha$ enumerates the sum over energy eigenstates. This expectation
value in Eq.~\ref{eq:SQP} is taken not only over the lattice ensemble, but over
the range of spatial paths that provide the same $r$ value. In this work, we
consider a maximum of $16$ on-axis points, and a range of $0$ to $3$ off-axis
points. The temporal extent considered has a maximum of $t=12$ for the untouched
and vortex-removed ensembles, and a maximum of $t=32$ for the vortex-only. The
larger value for the vortex-only ensemble is used because the onset of noise
occurs much later, and we find better plateau fits using this extended range.

Due to the cubic symmetry of the lattice, when considering a link path between two
spatial points separated by a given displacement vector $\vec{r} = \vec{y} - \vec{x}$ it is possible
to permute the three spatial coordinates and obtain the same value for the separation $r = |\vec{r}\:\!|$. Averaging over these permutations allows for further
improvement of statistics for the corresponding Wilson loop and better extraction of the ground state.

\section{Variational Analysis}\label{sec:VarAnalysis}

The analysis of the static quark potential is susceptible to excited state contamination and signal
to noise challenges. In particular, the dynamical ensembles are typically
noisier at a given lattice spacing compared to the pure gauge case. To better
extract the ground state potential at earlier Euclidean time, we create a
correlation matrix by introducing different levels of smearing along the two
spatial edges of the Wilson loops describing the profile of the flux tube,
\begin{equation}
  W_{ij}(r, \, t) = \Tr R_{i}(\vec{x},t_0)T(\vec{y},t_0)R_{j}^\dagger(\vec{x},t_1)T^\dagger(\vec{x},t_0).
\end{equation}
Here the forward and backwards paths $R_{i}(\vec{x},t_0)$ and $R_{j}^\dagger(\vec{x},t_1)$ are constructed using links that have respectively had $i$ and $j$ sweeps of spatial APE smearing~\cite{Albanese:1987ds} applied,
with a smearing parameter of $\alpha = 0.7$.
For the untouched and vortex-removed ensembles, the
$SU(3)$ projection component of the APE smearing algorithm is performed using
the unit-circle projection method described in Ref.~\cite{Kamleh:2004xk}.

The vortex-only ensembles present some difficulties in the application
of standard smearing algorithms, as highlighted by recent
work~\cite{Virgili:2022ybm} that delved into the difficult question of
smoothing $SU(3)$ centre vortex configurations. We employ these
findings to best extract the static quark potential, starting with a
brief summary of the relevant results from this study.

It was shown in Ref.~\cite{Virgili:2022ybm} that gauge-equivariant smoothing
(such as unit-circle projection) when applied to $SU(3)$ vortex-only
configurations results in either no effect or a swapping of the centre phase to
another element of $\mathbb{Z}_3$, spoiling the centre vortex structure. The use
of a non-analytic reuniterisation performed via a ${\rm Max}{\rm Re}\Tr$
method~\cite{Moran:2008ra} can circumvent this issue, however it is subject to
strict constraints on the smearing parameter $\alpha$.

The primary cause of the difficulties in smoothing vortex fields
arises from the proportionality of the vortex links to the
identity. To alleviate this issue, we apply the novel centrifuge
preconditioning method that was introduced in
Ref.~\cite{Virgili:2022ybm}, but only to the spatial links used to
construct the Wilson loop. Centrifuge preconditioning introduces a
small perturbation that rotates the vortex links away from the centre
group $\mathbb{Z}_3$ whilst maintaining the vortex structure. This is
then followed by application of APE smearing at smearing fraction
$\alpha_{\rm APE}=0.7$ using ${\rm Max}{\rm Re}\Tr$ reuniterisation to
generate the variational basis for vortex-only configurations.

For $N$ choices of smearing sweeps, we obtain the $N\times N$ correlation matrix
\begin{align}
  \label{eq:Gij}
  G_{ij}(r,t) &= \langle W_{ij}(r,t) \rangle \nonumber\\
            &= \sum_\alpha \lambda_i^{\alpha}\lambda_j^{*\alpha}\, \exp \left(-V^{\alpha}(r)\, t\right)
\end{align}
where the $i, \, j$ indices enumerate the $N$ smearing variations on the initial
and final spatial edges of the Wilson loop respectively. The complex scalars
$\lambda_i^{\alpha}$ and $\lambda_j^{*\alpha}$ represent the coupling of
each smeared leg of the Wilson loop to the static quark potential $V^{\alpha}$.
Note that in the following we choose to suppress the implied $r$ dependence of $G_{ij}$ and $V$ for clarity.

Presuming that the signal is dominated by the $N$ lowest energy states, such
that $\alpha \in [0, \, N - 1]$, we wish to find a basis ${\bf u}^{\alpha}$ such
that,
\begin{equation}
  \label{eq:basis1}
  G_{ij}(t)\, u_j^{\alpha} = \lambda_i^{\alpha}\, z^{*\alpha}\, e^{-V^{\alpha}\, t}\,,
\end{equation}
where $z^{*\alpha} = \sum_{i}\lambda_i^{*\alpha}\, u_i^{\alpha}$
is now the coupling between this new basis and the energy eigenstate
$|\alpha\rangle$. Note that for the remainder of this paper we adopt
the convention that repeated Latin indices are to be summed over whilst repeated
Greek indices are not. Eq.~(\ref{eq:basis1}) is equivalent to requiring that
\begin{equation}
  \lambda_i^{*\alpha}\, u_i^{\beta} = z^{*\alpha}\, \delta^{\alpha\beta}.
\end{equation}

Noting that the time dependence in Eq.~\ref{eq:basis1} depends only on the
exponential term, we can consider stepping forward in time by some amount
$\Delta t$ such that,
\begin{align}
  G_{ij}(t_0 + \Delta t)\, u_j^{\alpha} &= \lambda_i^{\alpha}\, z^{*\alpha}\, e^{-V^{\alpha}\, (t_0 + \Delta t)} \nonumber \\
                                        &= e^{-V^{\alpha} \Delta t}\, G_{ij}(t_0)\, u_j^{\alpha}\,.
\end{align}
This recursive relationship is precisely a generalised eigenvalue problem, which
can be solved via standard numerical techniques to obtain the eigenvectors
${\bf u}^{\alpha}$. An identical argument can be made for the left eigenvectors
${\bf v}^{\alpha}$, such that they satisfy
\begin{equation}
  v_i^{\alpha}\, G_{ij}(t)  = z^{\alpha}\, \lambda_j^{*\alpha}\, e^{-V^{\alpha}\, t}\,,
\end{equation}
and hence
\begin{equation}
  \label{eq:basis2}
  v_i^{\alpha}\, G_{ij}(t_0 + \Delta t) = e^{-V^{\alpha} \Delta t}\, v_i^{\alpha}\, G_{ij}(t_0)\, .
\end{equation}

Making use of Eq.~\ref{eq:basis1} and Eq.~\ref{eq:basis2}, we find that
\begin{equation}
  v_i^{\alpha}\, G_{ij}(t)\, u_j^{\beta} = z^{\alpha}\, z^{*\beta}\, \delta^{\alpha\beta}\, e^{-V^{\alpha}\, t}\, .
\end{equation}
As such, we define the eigenstate-projected correlator
\begin{align}
  G^{\alpha}(t) &= v_i^{\alpha}\, G_{ij}(t)\, u_j^{\alpha} \nonumber \\
                &= z^{\alpha}\, z^{*\alpha}\, e^{-V^{\alpha}\, t}\,,
\end{align}
and extract the potential by computing the log-ratio
\begin{equation}
  \label{eq:2}
  V_{\rm eff}^{\alpha}(t) = \ln \left( \frac{G^{\alpha}(t)}{G^{\alpha}(t + 1)} \right)\,,
\end{equation}
to obtain the static quark potential. We then consider constant fits to the lowest
energy state, $V_{\rm eff}^0(r,t)$.

We use a $4\times 4$ correlation matrix for the untouched and vortex-removed
ensembles, with a basis constructed from 6, 10, 18 and 30 sweeps of APE
smearing. For the vortex-only ensembles, even with centrifuge preconditioning
and ${\rm Max}{\rm Re}\Tr$ reuniterisation applied, the configurations are still slow to
vary as a function of  smearing sweeps. As a consequence of this, we choose a $2\times 2$ correlation matrix with 2
and 60 sweeps of APE smearing to provide a meaningful distinction between the
basis elements.

In regards to the choice of variational parameters for the original and
vortex-removed ensembles, we find that increasing $\Delta t$ minimally affects
the level of noise, whilst providing slight improvement in ground state
identification. Thus, we choose a larger value of $\Delta t = 3$. Selecting
larger values of $t_0$ introduces substantial noise into the results obtained
from these ensembles, so we maintain $t_0 = 1$ on these ensembles.

Selection of variational parameters is slightly different on the vortex-only
ensembles. For the diagonal correlators, $G_{ii}(t)$, where source and sink
match and all states should contribute positively, i.e.
$\lambda_i^{\alpha}\lambda_i^{*\alpha} > 0$, the effective mass approaches from
below. This is indicative of short-distance positivity violation arising in the
process of centre projection. In the context of a variational analysis, we
extend $t_0$ to the greatest feasible degree to avoid the region of positivity
violation at early times~\cite{Luscher:1984is}. Indeed, our focus is on
understanding whether projected centre vortices can capture the long-distance,
nonperturbative features of QCD. To this end, we choose
  $(t_0,\,\Delta t)$ to be $(5,\,4)$, $(4,\,5)$ and $(4,\,2)$ for the pure
gauge, $m_{\pi}=701~\si{MeV}$, and $m_{\pi}=156~\si{MeV}$ vortex-only ensembles
respectively. The difference in variational parameters between the ensembles
arises from when the onset of noise dominates the signal.

\begin{figure}
  \centering \subfloat{
    \includegraphics[width=0.9\linewidth]{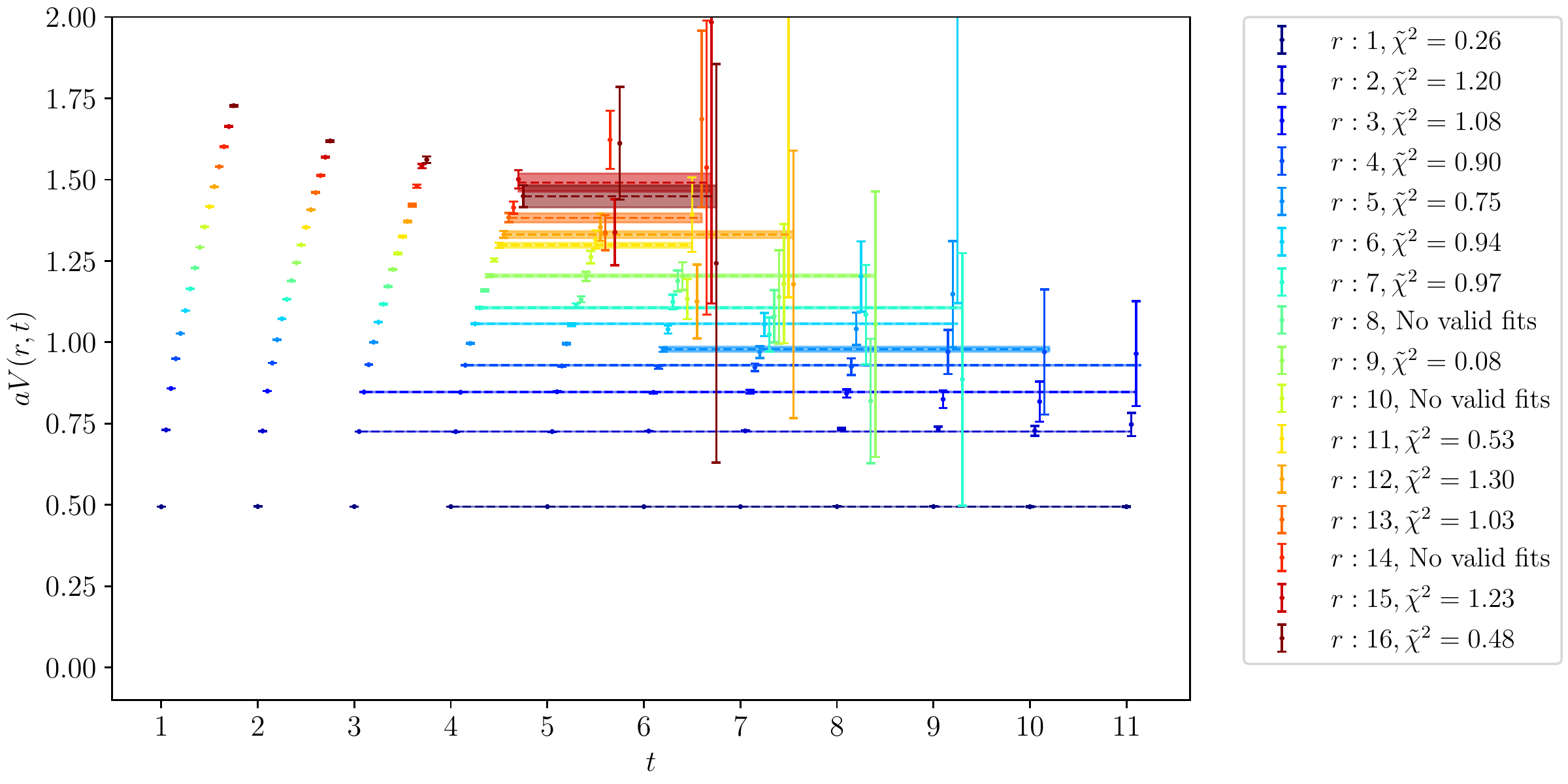}
    \label{fig:UT_E0}
  }%
  \newline \subfloat{
    \includegraphics[width=0.9\linewidth]{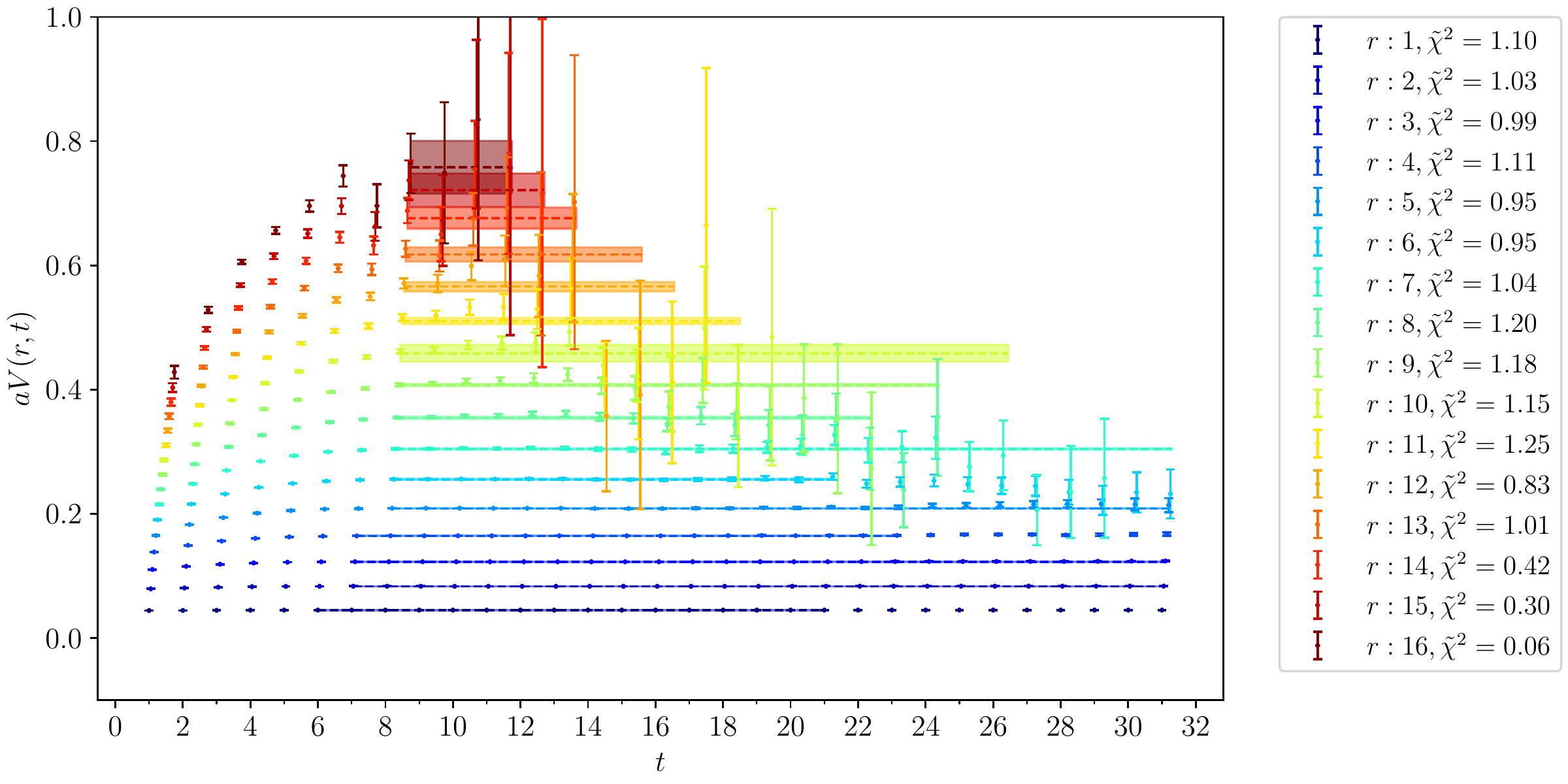}
    \label{fig:VO_E0}
  } \newline \subfloat{
    \includegraphics[width=0.9\linewidth]{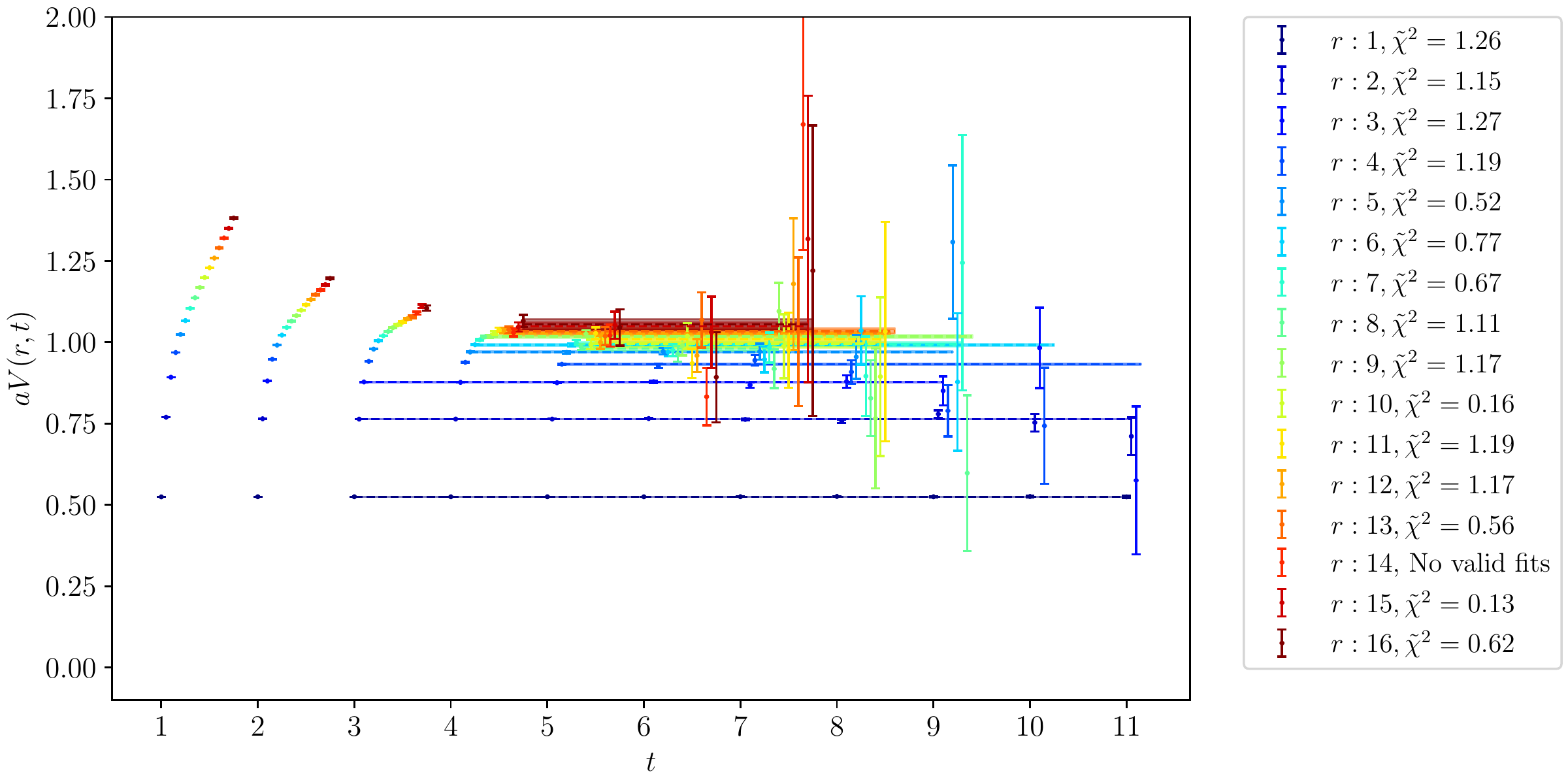}
    \label{fig:VR_E0}
  }
  \caption{\label{fig:E0all} The on-axis projected effective mass from the
    original $m_{\pi} = 156\, \si{MeV}$ ensemble. Results are shown for the
    original ({\bf top}), vortex-only ({\bf middle}) and vortex-removed ({\bf
      bottom}) ensembles. The selected fit window that meets the
    $\tilde{\chi}^{2}$ criteria as described in the text is shown as the dashed
    lines. The shaded region shows the jackknife error on the fit. Points
    at the same value of $t$ are horizontally offset for visual clarity.
    Any points with a relative error greater than 50\% are
      excluded from the plot.}
\end{figure}

To calculate uncertainties, we perform a third-order single-elimination
jackknife calculation~\cite{Leinweber:1990dv}. Fit window selection is performed
to prioritise finding the earliest appropriate value of $t_{\rm min}$, in a
method similar to that outlined in Ref.~\cite{Mahbub:2009nr}. As such, we select
an initial $t_{\rm max}$ to be the largest value maintaining
$V(r,t_{\rm max}) > \Delta V(r,t_{\rm max})$, where $\Delta V(r,t_{\rm max})$ is
the jackknife uncertainty in $V(r,t_{\rm max})$. An initial
  value of $t_{\rm min} = t_0 + 2$ is chosen. $t_{\rm max}$
  is then decreased until a covariance fit over the range
  $[t_{\rm min},\, t_{\rm max}]$ produces a $\chi^{2}$ per degree of freedom,
  $\tilde{\chi}^{2}$, of less than $1.3$. If no such $t_{\rm max}$ is found,
$t_{\rm min}$ is increased by one lattice unit and the procedure is repeated.
The on-axis results of this fitting procedure are shown for the lightest pion
mass ensemble in Fig~\ref{fig:E0all}. Once fits have been performed for all
values of $r$, we select a single fit window with a width of at least two
lattice units (i.e. at least three time values) such that it is typically
encompassed by the range of fit windows found for each value of $r$.

\begin{table}[tb]
  \caption{\label{tab:Ansatz}The ans\"atze used for the three ensembles.}
  \begin{ruledtabular}
    \begin{tabular}{lll}
      Type & Ansatz & Functional form\\
      \midrule
      Untouched & Cornell & $V(r) = V_0 - \alpha/r + \sigma\, r$\\
      Vortex-only & Linear & $V(r) = V_0 + \sigma\, r$\\
      Vortex-removed & Coulomb & $V(r) = V_0 - \alpha/r$
    \end{tabular}
  \end{ruledtabular}
\end{table}

After the potential $V(r)$ is determined, we then perform functional fits to the
UT, VO and VR potentials. The ans\"atze used for each ensemble are given in
Table~\ref{tab:Ansatz}. The functional fits take into account the full
covariance matrix, and error regions are constructed via repetition of the fits
on the jackknife ensembles. The selection of the range
$[r_{\rm min},\,r_{\rm max}]$ to fit over is performed in a manner similar to
the fit window selection for the effective mass. For the UT and VR ensembles we
initialise $r_{\rm min}$ to the lowest available value, as we find that our
window selection method naturally avoids the short-range region that is plagued
by lattice systematics. To explicitly avoid this region for the vortex-only
potential, we initialise $r_{\rm min} = 5$ for these ensembles. $r_{\rm max}$ is
initialised to the largest available value on all ensembles. Over this initial
range, the functional fit is performed and the $\chi^{2}$ per degree of freedom,
$\tilde{\chi}^{2}$, is calculated. If it is greater than $1.3$ then
$r_{\rm max}$ is reduced by $\Delta r=0.2$ and the fit is repeated. If
$r_{\rm max}-r_{\rm min}<3$, $r_{\rm max}$ is reset to its maximum extent and
$r_{\rm min}$ is increased by $\Delta r=0.2$. In our plots, points that are
included in the fit are shown in solid colours, whereas points excluded from the
fit are shown as faded.

We also present plots of the local slope calculated from a series of linear fits
taken over a sliding $r$ window of width 4 lattice units. Each fit window is
successively shifted in increments of $\Delta r = 0.4$ lattice units, with the
fitted slope plotted at the left-most edge. We find that $r = 5$ is
approximately where the onset of linearity begins, and hence we begin our
sliding windows from this value. The excluded short-distance region is greyed
out in the plots presented. This procedure for obtaining the local slope
provides a simple method for gauging the linearity of the potential over a range
of distances.

\section{Results}\label{sec:Results}

We now present the results for the static quark potential. To verify that our
variational technique is appropriate, we first calculate the vortex-only
potential from the $m_{\pi}=156~\si{MeV}$ ensemble without a variational method
to check if the results from the variational analysis are consistent and
represent an improvement. Given the similarity of the lattice spacing on our
three ensembles, summarised in Table~\ref{tab:LatticeParams}, we will consider
$r$ in lattice units for the remainder of this work. We find
  that the fitted string tension is lower after a variational analysis, with
  $\sigma_{\rm VO} = 0.0484(4)$ and $\sigma_{\rm VO} = 0.0490(4)$ with and
  without variational analysis respectively. Additionally, the effective mass
plateau fits occur at earlier times with the variational analysis, especially at
larger $r$ values. This suggests that the variational analysis is appropriate
and represents an improvement over the naive method.

\begin{figure}[t]
  \centerline{\includegraphics[width=0.9\linewidth]{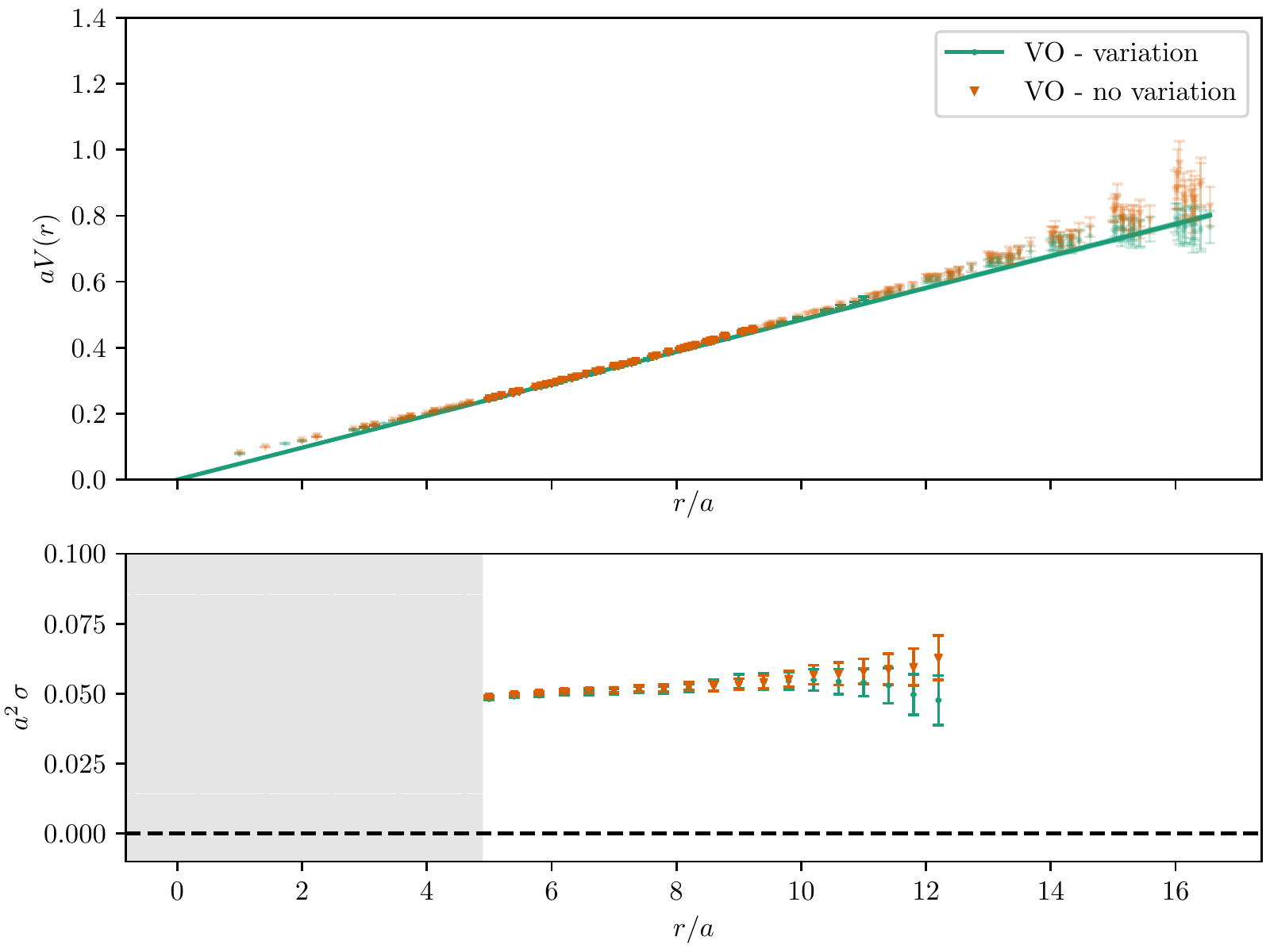}}
  \caption{\label{fig:VOcomp} A comparison of the vortex-only potential from the
    $m_{\pi} = 156 ~ \si{MeV}$ ensemble extracted after no spatial smearing and
    our variational method as described in the previous section. $V_0$ is set to
    $0$ for both sets of results. The functional fit for the variational results
    is also plotted. We observe a similar potential for both choices, however
    the linearity of the fit is improved after a variational method, with a
    larger range of points meeting the fit criteria discussed in the text.}
\end{figure}

We show the VO potential with and without variational analysis in
Fig.~\ref{fig:VOcomp}. Fitting is performed via the method outlined in the
previous section. We observe from the local slope plot that the long range
potential is similar across both methods. The fact that the differences are so
slight is a testament to the excellent signal-to-noise ratio in vortex only
ensembles and the subsequent access to large Euclidean times in the Wilson
loops. Nevertheless, the use of a variational method does improve the onset of
lower-lying plateaus and is thus preferred.

\subsection{Standard potential fits}

\begin{figure}[tbp]
  \centerline{\includegraphics[width=0.9\linewidth]{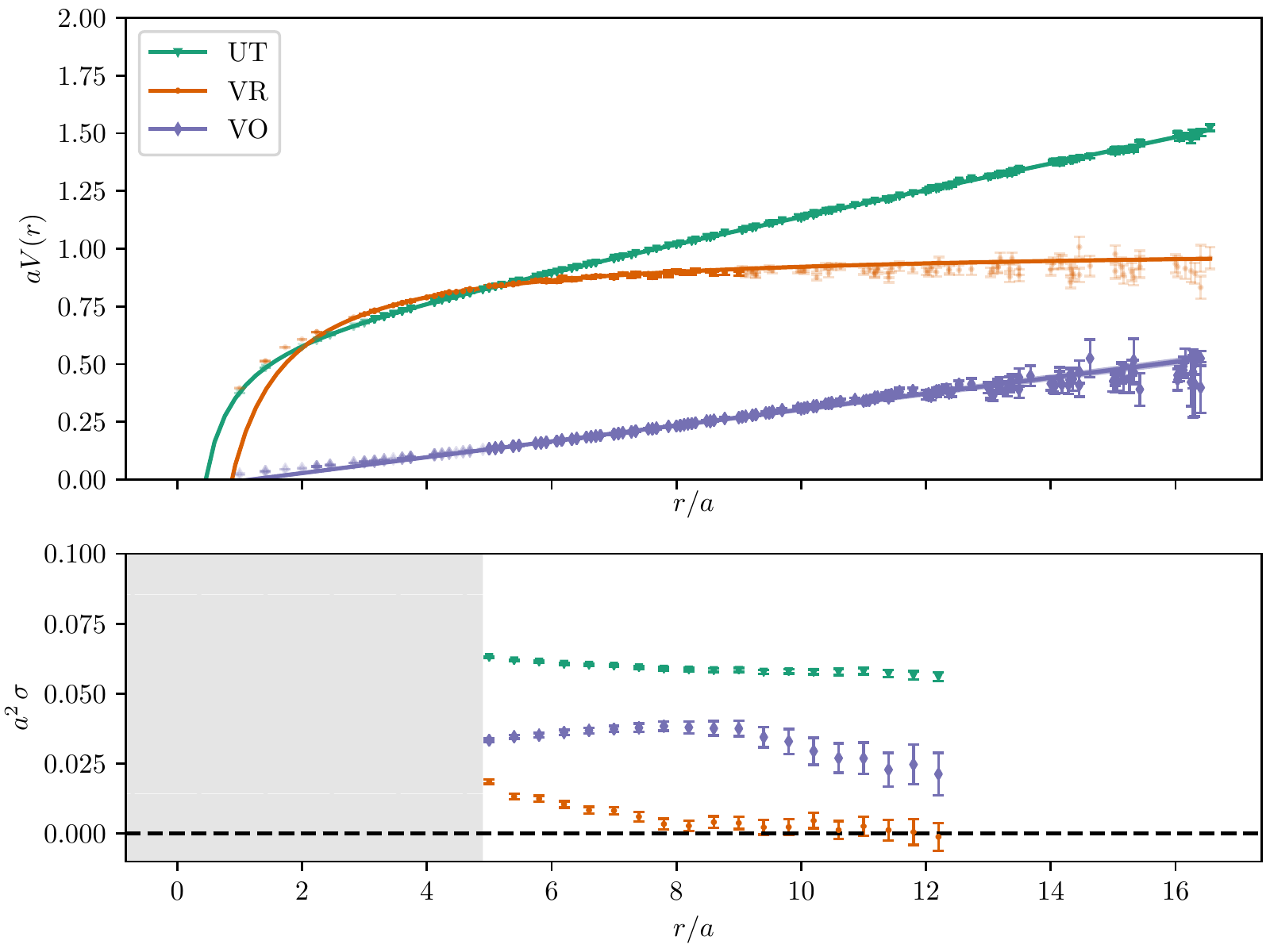}}
  \caption{\label{fig:SQPPG} The static quark potential as calculated from the
    pure Yang-Mills ensemble. Points are obtained from the variational analysis
    and solid lines show the fitted ansatz for each ensemble. The choice of
    ansatz is as described in Table~\ref{tab:Ansatz}. Faded points indicate that
    this point was not included in fitting the ansatz, as described in the text.
    The lower plot shows the fitted local slope of a forward-looking sliding
    linear window from $r$ to $r + 4a$.}
\end{figure}

The static quark potential from the pure gauge ensemble is presented in
Fig.~\ref{fig:SQPPG}. Our results coincide with findings from previous
studies~\cite{Langfeld:2003ev,OCais:2008kqh,Trewartha:2015ida}. The untouched
potential is Coulomb-like at short distances whilst becoming linear as $r$
increases. We observe that the vortex-removed and vortex-only potentials of
Table~\ref{tab:Ansatz} qualitatively capture these regimes respectively. Vortex
removal results in Coulomb-like behaviour at short distances, with approximately
constant behaviour at moderate to large $r$ indicating the absence of a linear
string tension. We do note, however, that the Coulomb term provides a poor
representation of the VR results at large $r$. Contrasting the vortex-removed
results, we observe that the vortex-only ensemble features no $1/r$ behaviour,
instead displaying a linear potential with a slope of approximately $62 \%$ that
of the original ensemble.

The fitted string tension values from the untouched and vortex-only ensembles
are presented in Table~\ref{tab:SQP}. The ratio of the vortex-only string
tension to the untouched string tension is shown in the third column. We see
that while the vortex field from the pure gauge background is only able to
recreate 62\% of the untouched string tension, in the presence of dynamical
fermions there is a different story. The fitted vortex-only string tension
increases upon the introduction of dynamical fermions at the heaviest pion mass.
At $m_\pi = 701~\si{MeV}$ the fitted string tension for the vortex-only and
untouched fields are nearly equal, whereas
on the lightest ensemble at
$m_\pi = 156~\si{MeV}$ the fitted string tension on the vortex-only field
exceeds the untouched value by about 25\%.

\begin{figure}[tbp]
  \centerline{\includegraphics[width=0.9\linewidth]{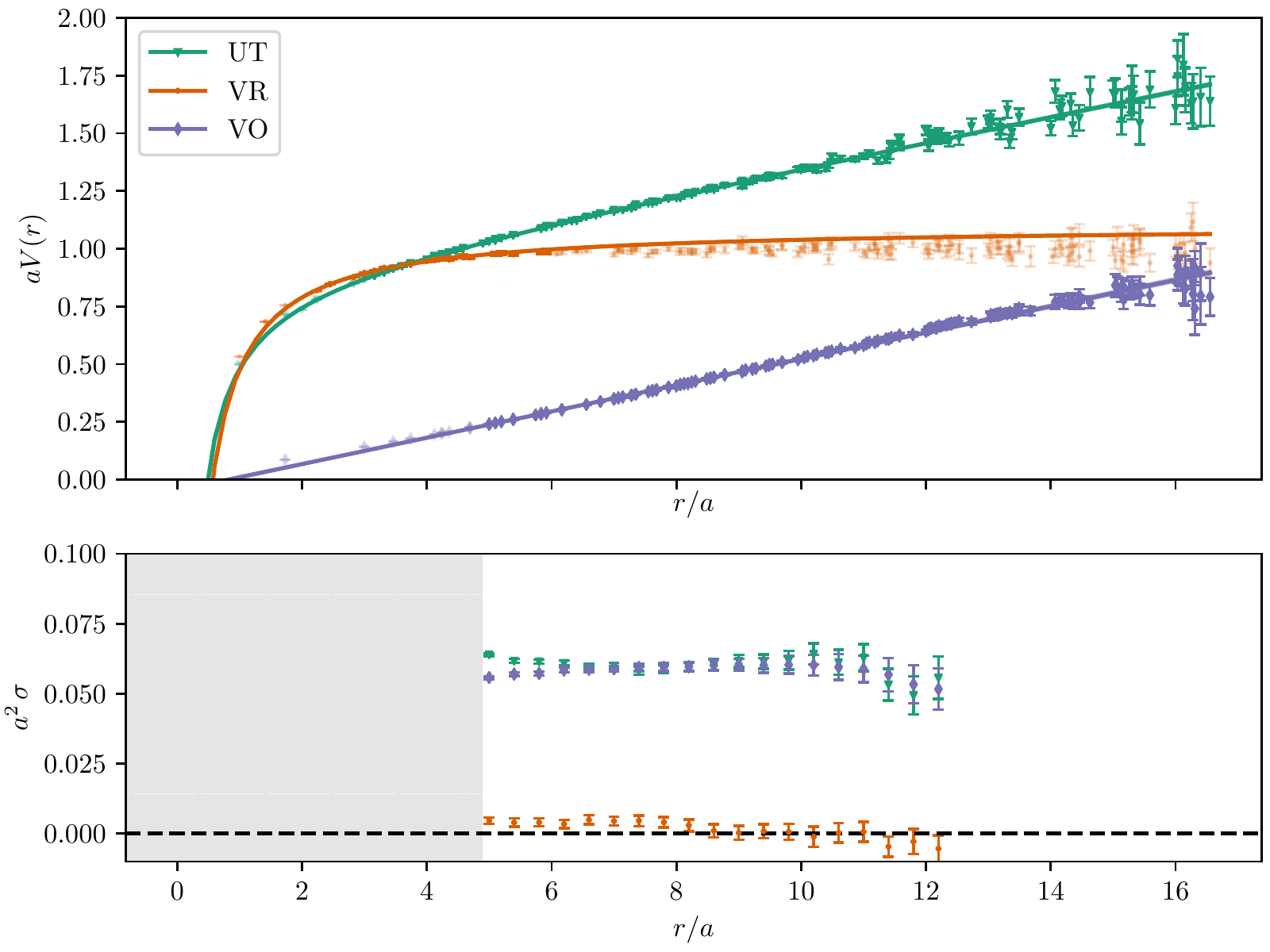}}
  \caption{\label{fig:SQPheavy} The static quark potential as calculated from
    the $m_{\pi} = 701 ~ \si{MeV}$ ensemble, with features as described in Fig.~\ref{fig:SQPPG}.}
\end{figure}

What is clear is that that the presence of dynamical fermions
significantly alters the texture of the vortex vacuum, even at an
unphysically large quark mass. The question then posed is how best to
shed some light on the nature of this `sea change.'
Fig.~\ref{fig:SQPheavy} shows the static quark potential results for the heavy dynamical ensembles, with
$m_{\pi} = 701 ~ \si{MeV}$.
Examining the local slope as it varies with $r$ provides some insight.
Note that the lattice spacings (as set by the S\"ommer scale) of the three ensembles listed in Table~\ref{tab:LatticeParams} are approximately the same, so it is reasonable to make broad comparisons in the slopes of the potentials.

%
\begin{figure}[tbp]
  \centerline{\includegraphics[width=0.9\linewidth]{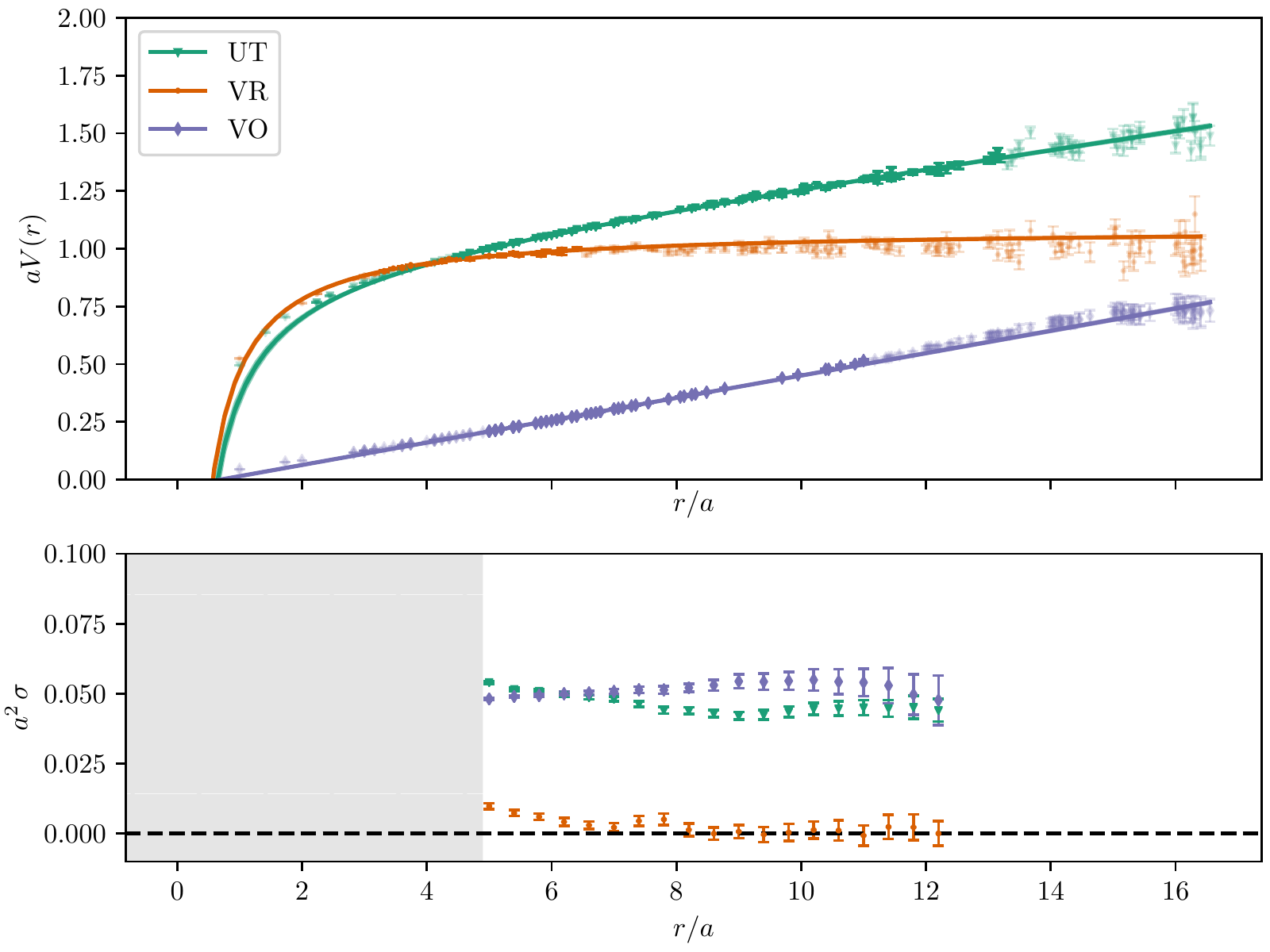}}
  \caption{\label{fig:SQPlight} The static quark potential as calculated from
    the $m_{\pi} = 156 ~ \si{MeV}$ ensemble, with features as described in Fig.~\ref{fig:SQPPG}.}
\end{figure}

As before, vortex removal captures the short-range physics while absenting any linear rise associated with a confining potential. Strikingly, the vortex-only field projected from the dynamical ensemble now
fully reproduces the long-range potential. This is best observed in the moving
local slope displayed in the lower panel of Fig.~\ref{fig:SQPheavy}. The more precise fitted string tension $\sigma$ shows approximate agreement
as reported in Table~\ref{tab:SQP}. This will be discussed in greater detail in the next subsection.

\begin{table}[tb]
  \caption{\label{tab:SQP}The fitted string tensions from the
    vortex-only and untouched ensembles, and their respective ratios.}
  \begin{ruledtabular}
    \begin{tabular}{lddd}
      $m_{\pi}\, (\si{MeV})$ & \mc{a^2\,\sigma_{\rm VO}^{\rm }} & \mc{a^2\,\sigma_{\rm UT}^{\rm }} & \mc{\sigma_{\rm VO}/\sigma_{\rm UT}^{\rm }} \\
\midrule
Pure gauge & 0.0344(9) & 0.0558(3) & 0.62(2) \\
$701$ & 0.0570(7) & 0.0537(7) & 1.06(2) \\
$156$ & 0.0484(4) & 0.0386(1) & 1.25(3) 

    \end{tabular}
  \end{ruledtabular}
\end{table}

Finally, we present the static quark potential on the ensemble with the lightest
pion mass of $156~\si{MeV}$ in Fig.~\ref{fig:SQPlight}. Here we observe the
untouched and vortex-only slopes cross-over, with approximate agreement of the
local slope in the region $r \in [5.5, 7]$. As we extend to larger distances, we
observe that the vortex-only string tension exceeds the original value. This
overestimation is corroborated by the fit values, where the value of $\sigma$
reported in Table~\ref{tab:SQP} is approximately $25\%$ larger than the
untouched.
\begin{table*}[bt]
  \caption{\label{tab:allfit}Results of the standard static quark potential fits to the three ensembles. The fit parameters are described in Table~\ref{tab:Ansatz} and $\tilde{\chi}^{2}$ denotes the $\chi^{2}$ per degree of freedom.}
  \begin{ruledtabular}
    \begin{tabular}{lldddd}
      Type & \mc{(r_{\rm min},\, r_{\rm max})} & \mc{\tilde{\chi}^2} & \mc{aV_0} & \mc{\alpha} & \mc{a^2\sigma} \\
\midrule
\multicolumn{6}{l}{Pure gauge}\\
\midrule
UT & (3.10, 16.55) & 1.12 & 0.608(3) & 0.286(7) & 0.0558(3) \\
VR & (3.00, \phantom{0}9.05) & 1.23 & 1.010(2) & 0.881(7) & \cdash \\
VO & (5.00, 16.40) & 0.97 & -0.041(4) & \cdash & 0.0344(9) \\
\midrule
\multicolumn{6}{l}{$m_{\pi} = 701~\si{MeV}$}\\
\midrule
UT & (3.10, 16.55) & 1.30 & 0.847(7) & 0.42(1) & 0.0537(7) \\
VR & (3.00, \phantom{0}6.55) & 1.30 & 1.092(4) & 0.59(1) & \cdash \\
VO & (5.00, 16.55) & 1.03 & -0.047(4) & \cdash & 0.0570(7) \\
\midrule
\multicolumn{6}{l}{$m_{\pi} = 156~\si{MeV}$}\\
\midrule
UT & (4.40, 13.25) & 1.29 & 0.93(1) & 0.61(4) & 0.0386(1) \\
VR & (3.10, \phantom{0}5.40) & 1.28 & 1.106(5) & 0.68(2) & \cdash \\
VO & (5.00, 11.15) & 1.28 & -0.033(2) & \cdash & 0.0484(4) 

    \end{tabular}
  \end{ruledtabular}
\end{table*}


%
The unanticipated overestimation of the VO string tension at the lightest mass gives an indication that there is some additional physics that is not being accounted for. A hint as to the possible answer is revealed in the vortex-removed fits. Specifically, the standard Coulomb term retains a residual increase in strength at moderate to large $r$ that does not match the approximately constant behaviour of the vortex-removed results. The slow rise present in the standard Coloumb term could also interfere with the fitted linear term coefficient, resulting in an underestimation of the string tension in the UT results where both the Coulomb and string-tension terms are present.

Table~\ref{tab:allfit} shows that as pion mass decreases, the fitted value of
the Coulomb term coefficient, $\alpha$, on the UT ensembles increases. This
would then enhance possible contamination of the fitted UT string tension
resulting from physics absent from the standard Coulomb term, amplifying the
discrepancy between the original and vortex-only string tensions. This motivates
modifications to the Coulomb term that we introduce in the next section in order
to obtain better descriptions of the lattice results and more accurate estimates
of the string tension.


\subsection{Modified Coulomb potential fits}



We have seen the difficulty in fitting the Coulomb term 
parameter, $\alpha$, in our ansatz to a wide range of values on the
dynamical ensembles. At the shortest distances, there is a well-known difficulty
associated with fitting $\alpha$ for both the original and vortex-removed
ensembles~\cite{Aoki:1999ff}, stemming from the small statistical errors present
at short range coupled with the presence of finite lattice-spacing systematics.

It is possible to apply a lattice correction to the Coulomb term to
compensate for these short-distance artifacts~\cite{Michael:1992nj,Edwards:1997xf}.
However, here we are mainly concerned with the long distance behaviour and adopt the simple
solution of excluding small values of the static quark separation $r$
from our fits.

A more serious limitation in the fit functions used above is revealed
upon vortex removal. The standard Coulomb term is only able to
describe the vortex-removed results over a limited range. This demonstrates
a need for a modified fit function in order to describe the large $r$ behaviour of the vortex-removed potential.

The decoupling of the static quark potential into the vortex-removed
and vortex-only components also provides us with an
opportunity. Specifically, the large $r$ behaviour of the untouched
potential is dominated by the linear string tension. The dominance of
the linear term at large $r$ hides any subleading effects.

The vortex-only component of the potential is well described by a linear
string tension. The origin of the confining string tension is attributed to non-trivial vacuum structure, with the centre-vortex model of course being the most pertinent to this study.
On the other hand, the vortex-removed potential does
not possess a string tension as testified by the absence of a linear slope. This provides us with a chance to model
effects that would otherwise be obscured by the rising linear string tension.

The first modified ansatz we propose is novel, with a model based on anti-screening of the Coulomb potential,
\begin{equation}
  V_{\rm as} (r) = V_0 - \frac{\alpha}{ 1-e^{-\rho r}} \label{eq:anti-screen}\,.
\end{equation}
The Laurent series of this function is dominated by the lowest order term $\tilde{\alpha}/r$ at short distances providing a Coloumb-like potential, where the effective Coulomb coefficient is $\tilde{\alpha} = \alpha/\rho.$ Anti-screening implies that the strong coupling constant $\alpha_s(r)$ increases with increasing separation between two test colour charges. If $\alpha_s$ increases as $r$ increases, this will have the effect of counteracting decreasing behaviour of the $1/r$ term.

The specific form of the ansatz we have chosen here is motivated by the observation of the flat, constant-like behaviour of the vortex-removed potential at large distances. Specifically, at large $r$ the exponential in the denominator of Eq.~(\ref{eq:anti-screen}) tends to zero, such that a constant value $V_{\rm as} \to V_0 - \alpha$ is rapidly approached as $r$ increases. The implication of this is that the running coupling of $\alpha_s$ is approximately linear in $r$ within the fitted region. Previous lattice studies of the running of the strong coupling do show an increase in $\alpha_s$ with the separation $r,$ although they are limited in the applicable range of scale (up to $\sim$0.5 fm)~\cite{Michael:1992nj,Sommer:1993ce,Klassen:1994mw}. Importantly, the form of Eq.~(\ref{eq:anti-screen}) is controlled such that the large $r$ behaviour cannot describe a rising linear potential tension and hence should not interfere with a fitted string tension.

Intuitively, anti-screening can be understood by noting that at short distances
gluons carry colour charge away from a quark or anti-quark such that the
effective colour charge within a given radius is diluted, leading to asymptotic
freedom at short distances~\cite{Deur:2016tte}. We know from previous studies of
the pure-gauge vortex-removed gluon propagator that flat behaviour consistent
with asymptotic freedom is observed at large $q^2$~\cite{Biddle:2018dtc}. We
also know that anti-screening arises from the non-Abelian nature of the gluon
field, and as the vortex-removed field remains non-Abelian it seems reasonable
to postulate that anti-screening will still be present in the absence of
confinement.

\begin{table*}[t]
  \caption{\label{tab:parameters-extra}Results of the functional fits to the modified
    ans\"atze described in the text. The values of $\rho$ for the untouched
    ensembles are fixed to the value obtained from the corresponding
    vortex-removed fit.}
  \begin{ruledtabular}
    \begin{tabular}{lllddddd}
      Type & \mc{(r_{\rm min},\, r_{\rm max})} & Fit function & \mc{\tilde{\chi}^2} & \mc{a\,V_0} & \mc{\alpha} & \mc{a^2\,\sigma} & \mc{\rho} \\
\midrule
\multicolumn{6}{l}{Pure gauge}\\
\midrule
VR & (2.90, 16.55) & $V_{\rm as}$ & 1.10 & 1.20(3) & 0.27(3) & \cdash & 0.28(2) \\
VR & (2.90, 16.55) & $V_{\rm sc}$ & 1.13 & 0.931(5) & 1.01(3) & \cdash & 0.15(2) \\
UT & (3.00, 16.55) & $V_{\rm as} + \sigma\, r$ & 1.16 & 0.652(4) & 0.081(2) & 0.0572(3) & 0.28 \\
UT & (3.00, 16.55) & $V_{\rm sc} + \sigma\, r$ & 1.19 & 0.573(2) & 0.301(7) & 0.0572(3) & 0.15 \\
\midrule
\multicolumn{6}{l}{$m_{\pi} = 701~\si{MeV}$}\\
\midrule
VR & (1.80, 16.55) & $V_{\rm as}$ & 0.97 & 1.42(2) & 0.42(3) & \cdash & 0.53(2) \\
VR & (1.80, 16.55) & $V_{\rm sc}$ & 1.01 & 1.005(2) & 0.85(2) & \cdash & 0.31(2) \\
UT & (3.00, 16.55) & $V_{\rm as} + \sigma\, r$ & 1.29 & 1.02(1) & 0.259(9) & 0.0588(5) & 0.53 \\
UT & (3.00, 16.55) & $V_{\rm sc} + \sigma\, r$ & 1.30 & 0.761(4) & 0.54(2) & 0.0585(5) & 0.31 \\
\midrule
\multicolumn{6}{l}{$m_{\pi} = 156~\si{MeV}$}\\
\midrule
VR & (3.00, 16.40) & $V_{\rm as}$ & 1.18 & 1.48(6) & 0.48(6) & \cdash & 0.51(4) \\
VR & (3.00, 16.40) & $V_{\rm sc}$ & 1.18 & 1.009(3) & 1.05(8) & \cdash & 0.33(3) \\
UT & (4.40, \phantom{0}9.25) & $V_{\rm as} + \sigma\, r$ & 1.28 & 1.17(4) & 0.37(3) & 0.0459(9) & 0.51 \\
UT & (4.40, \phantom{0}9.25) & $V_{\rm sc} + \sigma\, r$ & 1.28 & 0.804(7) & 0.84(7) & 0.0457(9) & 0.33 

    \end{tabular}
  \end{ruledtabular}
\end{table*}
Of course there are more sophisticated calculations of the running of
$\alpha_s$~\cite{Booth:1992bm,Bali:1992ru,Luscher:1993gh,Klassen:1994mw,Blum:1994zf,Hornbostel:2002af,Bazavov:2014soa},
but these have limited applicability here, either due to the limited range of
perturbation theory in QCD or being inspired by the string tension. It is not
clear how these apply to vortex-modified fields. Here we choose instead to
simply model the observed behaviour of the vortex-removed potential.


We also consider an alternative model to fit the vortex-removed results. The second
modified ansatz we propose is a screened Coulomb potential, commonly known as
the Yukawa potential,
\begin{align}
  V_{\rm sc}(r) &= V_0-\frac{\alpha}{r}\,e^{-\rho r}\label{eq:yukawa}\,.
\end{align}
Once again this has a Coulomb-like $1/r$ behaviour at small $r.$ At large $r$
the exponential term has the effect of turning off the Coulomb interaction such
that $V_{\rm sc} \to V_0$ as $r$ increases.


One interpretation of the Yukawa model in this context is that the gluon
dynamically acquires an effective mass $\rho$ in the infrared. As a non-zero
gluon mass is forbidden at the Lagrangian level by gauge invariance, this
mechanism must be dynamical and scale-dependent. Indeed, the dynamical
generation of an effective gluon mass has been proposed elsewehere as a possible
mechanism for the gluon propagator to take a finite value in the infrared
limit~\cite{Mandula:1987rh,Bogolubsky:2007ud,Bogolubsky:2009dc,Li:2019hyv,Horak:2022aqx}.

It must be emphasised that the finiteness of the gluon propagator in the infrared limit is distinct to the presence (or absence) of confinement. The signature of confinement is dependent on the nature of the running of the gluon mass. Specifically, confinement is associated with an inflection point or turn-over in the gluon propagator, which in turn implies the running gluon mass should not be constant. We know that vortex-removed theory does not generate a string tension and hence is non-confining. Introducing the possibility of a constant effective gluon mass at a finite scale would model the vortex-removed potential in a way which is separate to any confinement mechanism.
\begin{figure*}[t]
  \centering
\begin{tabular}{cc}
  \subfloat[Pure gauge, $V_{\rm as}$ fit.]{
    \includegraphics[width=0.45\linewidth]{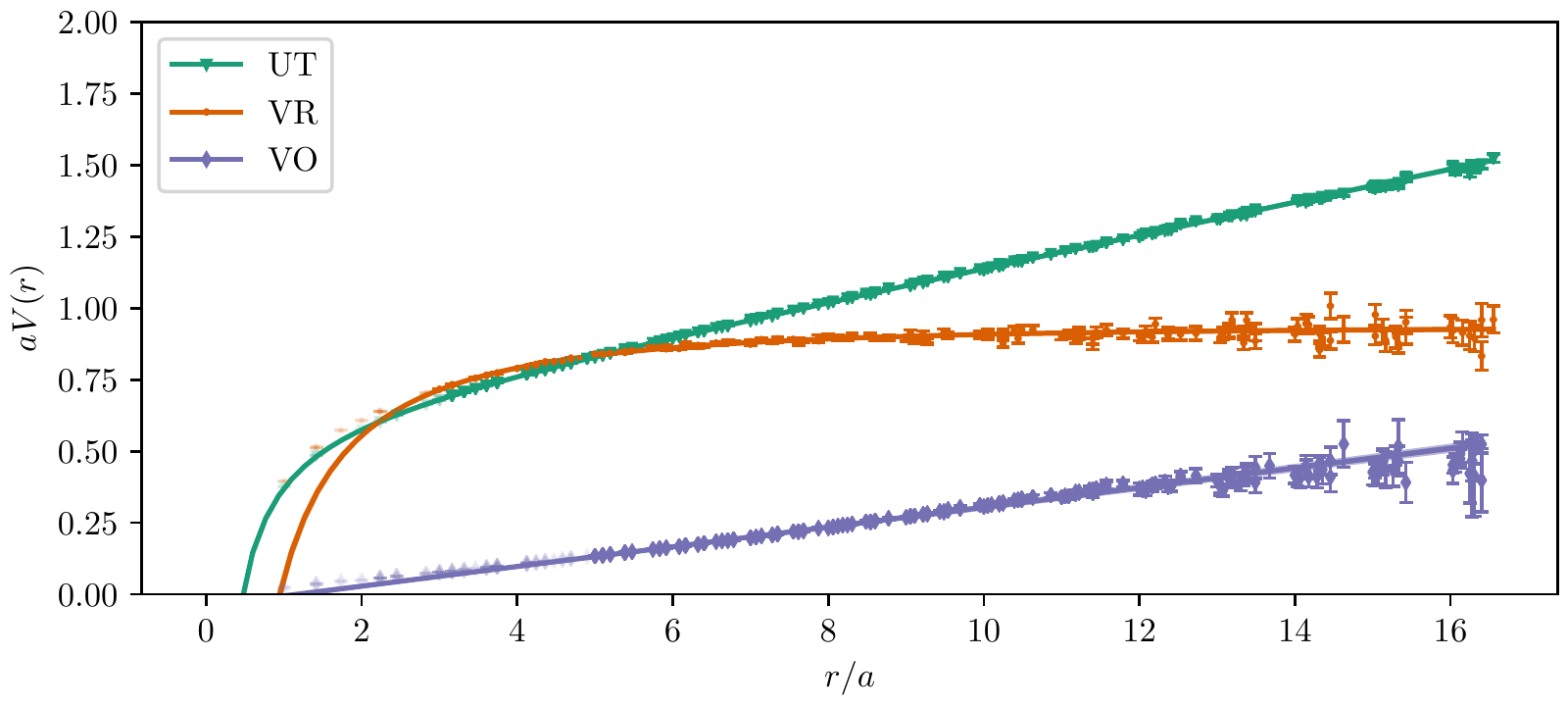}
  }  &
  \subfloat[Pure gauge, $V_{\rm sc}$ fit.]{
    \includegraphics[width=0.45\linewidth]{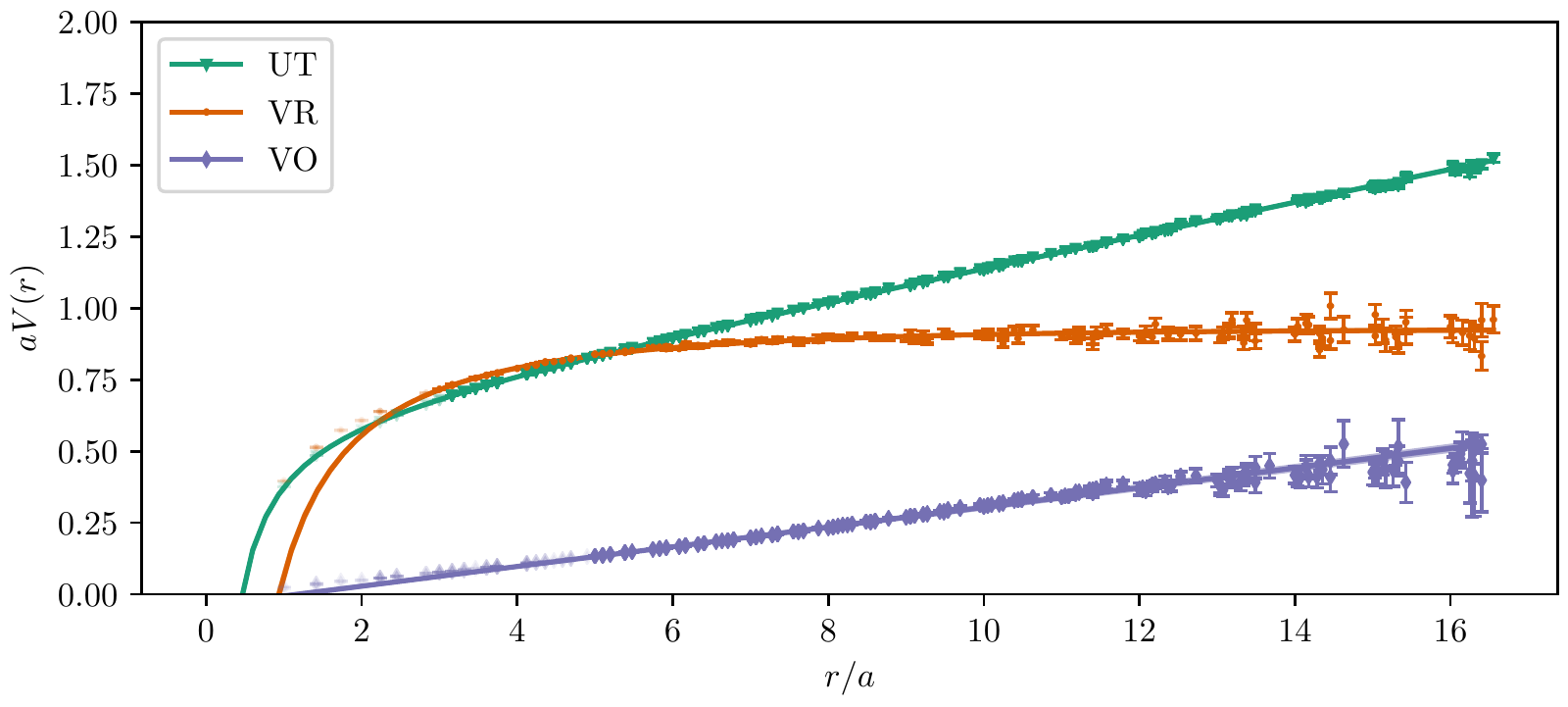}
  }\\
  \subfloat[$m_{\pi}=701~\si{MeV}$, $V_{\rm as}$ fit.]{
    \includegraphics[width=0.45\linewidth]{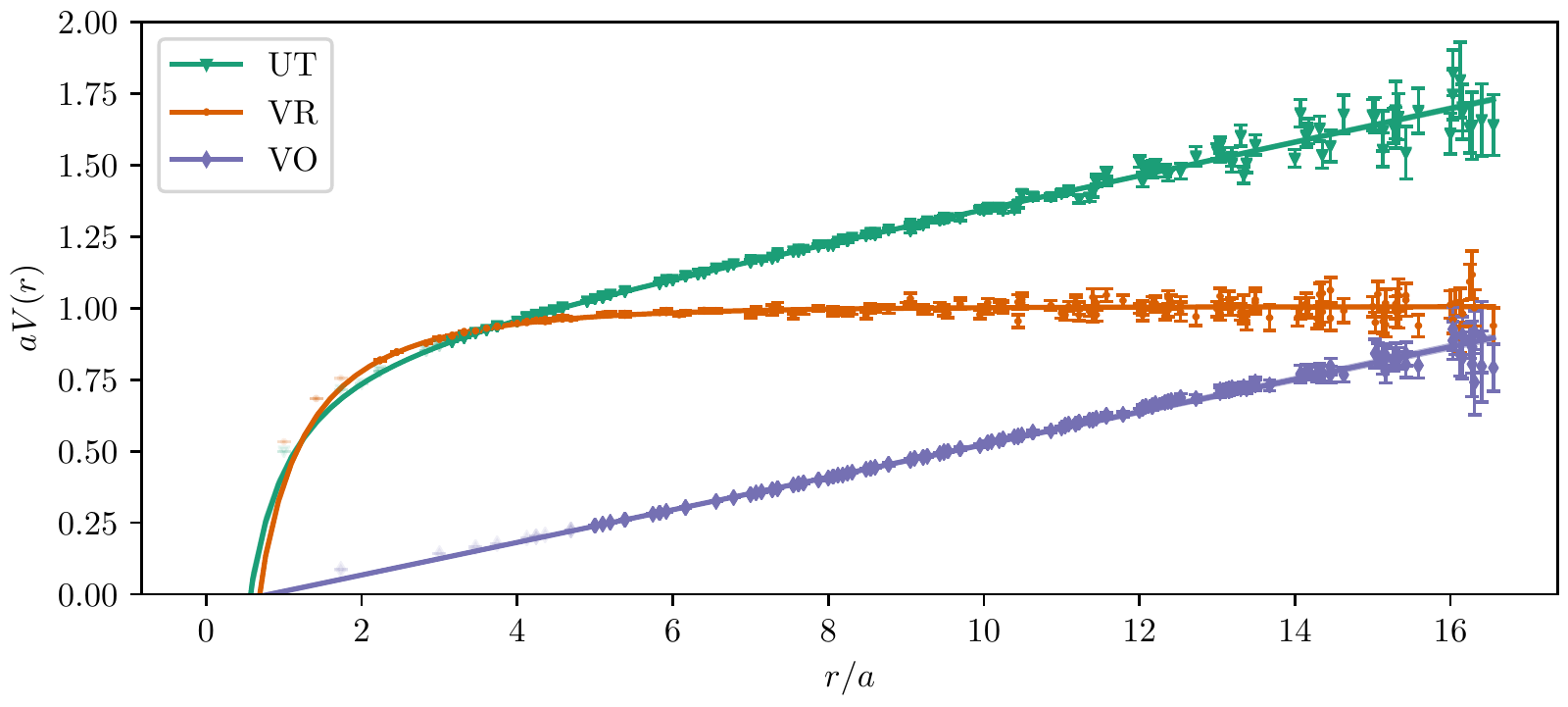}
  } &
  \subfloat[$m_{\pi}=701~\si{MeV}$, $V_{\rm sc}$ fit.]{
    \includegraphics[width=0.45\linewidth]{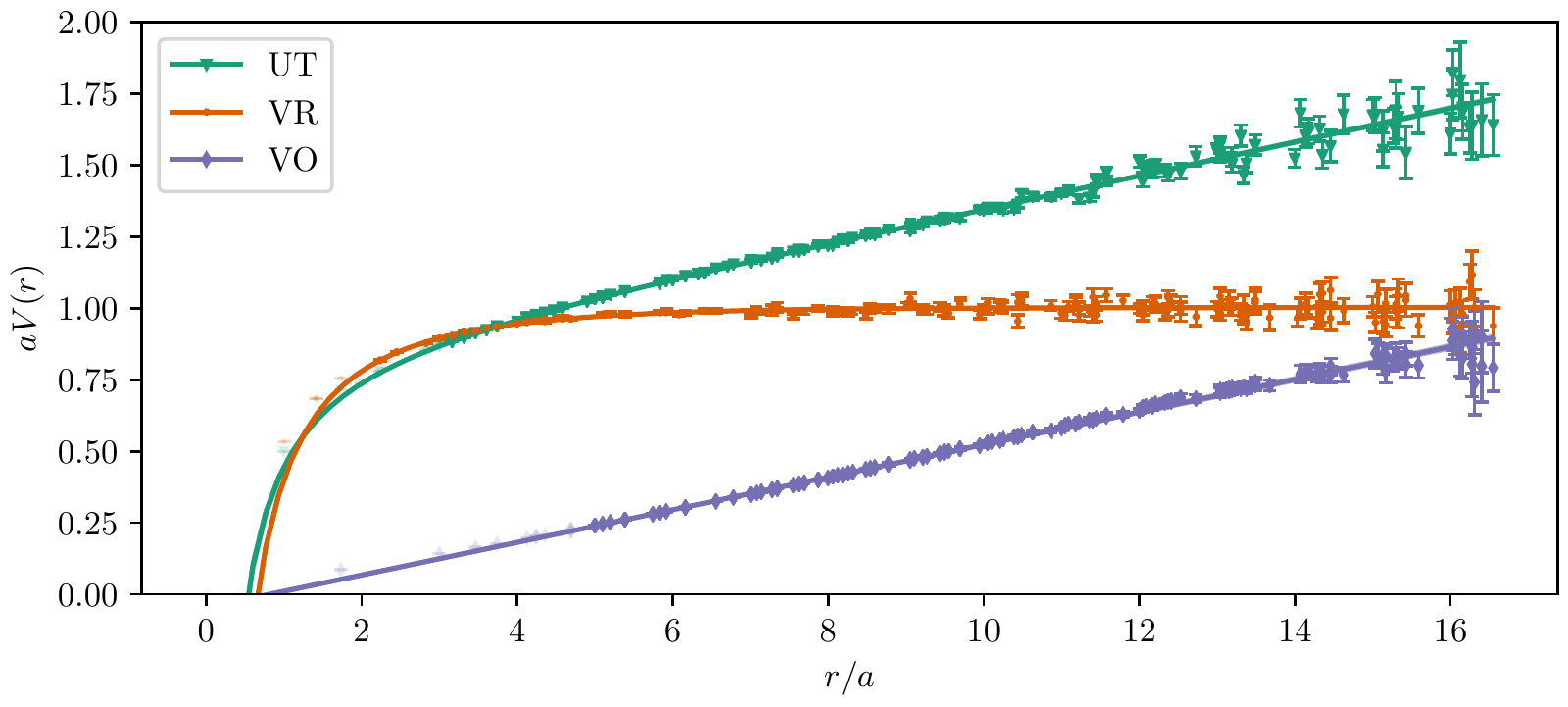}
  }\\
  \subfloat[$m_{\pi}=156~\si{MeV}$, $V_{\rm as}$ fit.]{
    \includegraphics[width=0.45\linewidth]{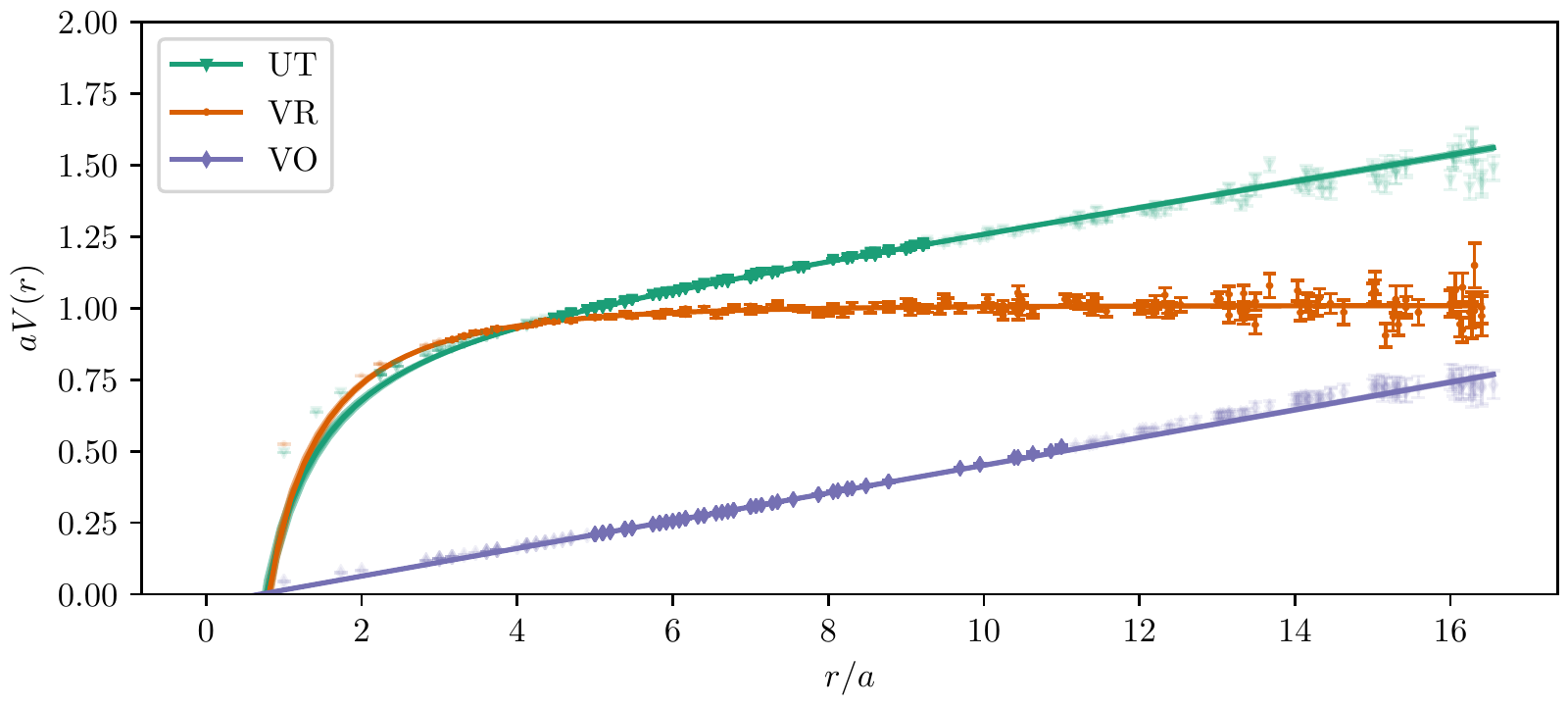}
  } &
  \subfloat[$m_{\pi}=156~\si{MeV}$, $V_{\rm sc}$ fit.]{
    \includegraphics[width=0.45\linewidth]{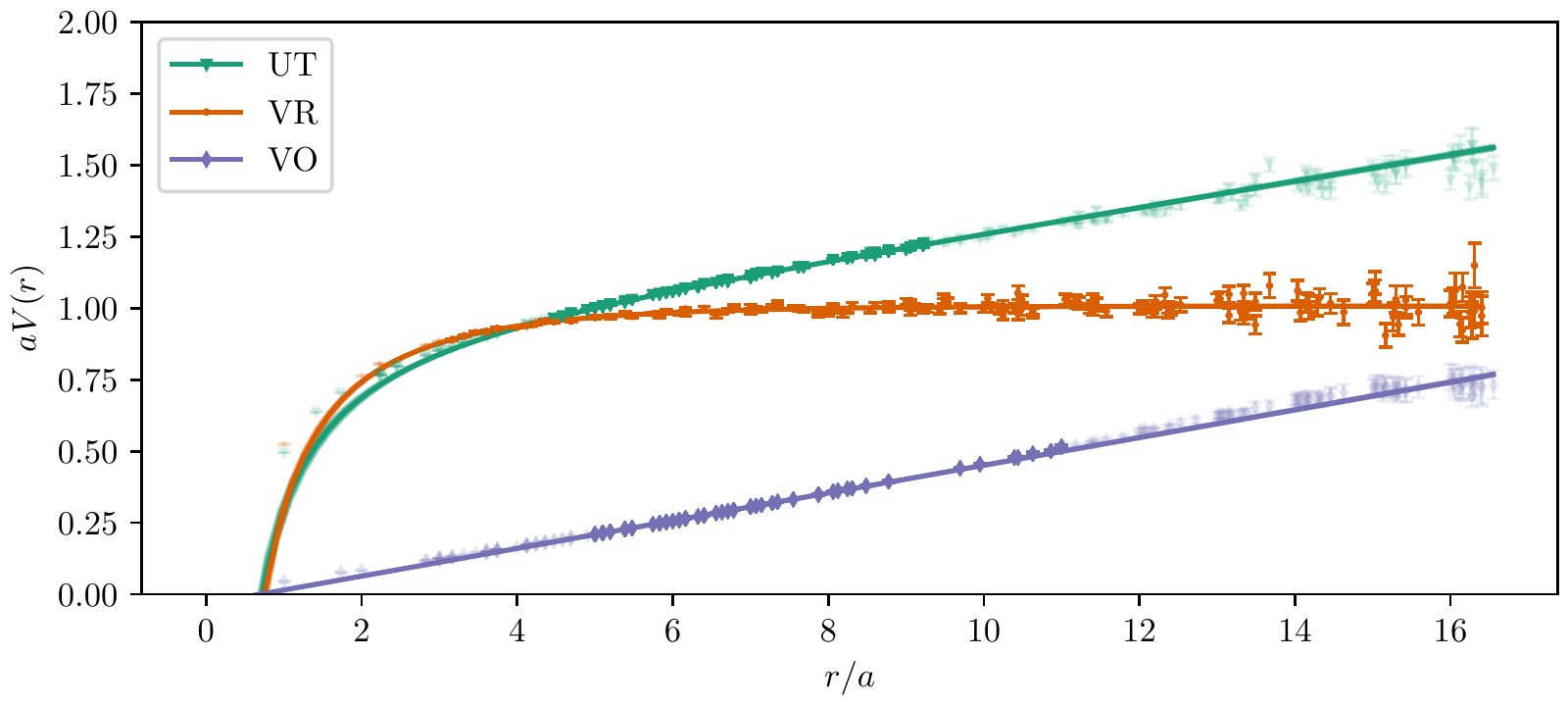}
  }
\end{tabular}

\caption{\label{fig:extra-plots-all} Fits to the lattice results for the
  potentials using the modified Coulomb term functions $V_{\rm sc}$ and
  $V_{\rm as}$ described in the text. The vortex-removed results are now
  described well by the modified potentials.}
\end{figure*}

We now turn to the results from our modified Coulomb ans\"atze. Table~\ref{tab:parameters-extra} presents the fit parameters, with the resulting potentials illustrated in Fig.~\ref{fig:extra-plots-all}. We see that both $V_{\rm as}$ and $V_{\rm sc}$ are able to describe the vortex-removed results well, with similar values for the reduced $\chi^2.$ At first glance it seems somewhat counter-intuitive that both an anti-screened and screened model are able to describe the same results. Numerically, this is possible because of the interplay between the $V_0$ and $\alpha.$ Both ans\"atze approach a constant value in the large $r$ limit, with $V_{\rm as} \to V_0 - \alpha$ and $V_{\rm sc} \to V_0$ respectively.

We see that both modified ans\"atze provide a superior fit to the vortex-removed results when compared to the standard Coulomb ansatz, allowing the fit window to extend to the maximum available $r_{\rm max}.$ In all cases the fitted value of $r_{\rm min}$ is less than or equal to the standard potential fits, indicating that the modifications made to the Coulomb terms  are still able to account for the short distance behaviour of the potential up to the presence of lattice artefacts.

Having verified that our modified ans\"atze are successfully able to describe the vortex-removed potential results at large $r,$ we can then use this information to improve our fits to the untouched results. This is accomplished by fixing $\rho$ to be the value obtained from the
corresponding vortex-removed ensemble, then adding a linear term to accommodate the string tension component of the untouched potential. The motivation behind fixing $\rho$ is that the cleanest fit value for this parameter will be obtained in the absence of a string tension term which will dominate the large $r$ behaviour. Indeed, we find that if left as a free parameter $\rho$ is poorly constrained by the untouched potential fits due to the presence of the dominating linear term.

The fits to the untouched ensembles are of comparable range and $\tilde{\chi}^2$
to the original Cornell fits, however when we look at the ratio of the
vortex-only string tension to the untouched, shown in Table~\ref{tab:SQP-extra},
we see the significant impact the modified Coulomb terms play. The
untouched string tension on the pure gauge ensemble is similar to the Cornell fit value,
however on the dynamical ensembles the string tension is increased due to cleanly removing the
contamination from the slow rise in the standard Coulomb term at moderate to large $r.$ Remarkably, this results in agreement between the vortex-only
and untouched string tensions on both dynamical lattices, as seen by the corresponding ratios taking values close to unity in Table~\ref{tab:SQP-extra}.

\begin{table}[tb]
  \caption{\label{tab:SQP-extra}Ratios of the vortex-only to untouched string tensions
    from the Cornell and modified fit functions.}
  \begin{ruledtabular}
    \begin{tabular}{ldddd}
      $m_{\pi}\, (\si{MeV})$ & \mc{\sigma_{\rm VO}/\sigma_{\rm UT}^{\rm cornell}} & \mc{\sigma_{\rm VO}/\sigma_{\rm UT}^{\rm as}} & \mc{\sigma_{\rm VO}/\sigma_{\rm UT}^{\rm sc}} \\
\midrule
Pure gauge & 0.62(2) & 0.60(2) & 0.60(2) \\
$701$ & 1.06(2) & 0.97(2) & 0.97(2) \\
$156$ & 1.25(3) & 1.05(2) & 1.06(2)

    \end{tabular}
  \end{ruledtabular}
\end{table}
The fits to the results are unable to distinguish between the two modified
ans\"atze. Indeed, the resulting improvements to the untouched potential fits
result in values for the string tension that are essentially identical. We also
tested an $n$-tuple form factor, $(1 + (r/\rho)^{n})^{-1}$, to suppress the
Coulomb term at large $r$, and this provided a similar result. This gives us
confidence that any systematic errors arising from the modified Coulomb terms
are minimal in the final string tensions reported.

The physical arguments provided for the two modified ans\"atze are simply to
demonstrate some plausible mechanisms that might underpin their empirically
motivated forms. Due to the interplay between $\alpha$ and $V_0$ it is likely
that more than one effect will contribute to the fitted values. With a
high-precision scaling analysis, a future examination may be able to resolve the
physics represented by these modifications. The key result here is that by
successfully modelling the observed long distance behaviour of the
vortex-removed potential, we have been able to remove a source of contamination
in the untouched potential fits and provide improved values for the fitted
string tension for the first time.

For a given ansatz, the fitted value of $\rho$ on the two dynamical lattices are
similar, and are roughly double the fit value on the pure gauge ensemble. This
indicates that the effects contributing to the medium to long-range behaviour
of the vortex-removed potential are mainly sensitive to the presence or absence
of dynamical fermions, but are only weakly dependent on the sea quark mass.

There are indications of increased screening by the light dynamical fermions in
both the untouched and vortex-only results. Significantly, at longer distances
we observe both modified ans\"atze show a decrease in the fitted value of the
untouched and vortex-only string tensions when transitioning from the heavy to
light pion mass.

As we have not corrected for short-distance lattice artefacts the fitted values
of $\alpha$ should be interpreted with some caution, but are also worth
discussing. The Coulomb term coefficients arising from the fits to the untouched
potentials are summarised in Table~\ref{tab:SQP-alpha} (recalling that for the
$V_{\rm as}$ ansatz the effective short-distance coupling is
$\tilde{\alpha} = \alpha/\rho).$ For the pure gauge ensemble, the fitted values
are close to the universal value of $\pi/12 \simeq 0.26$ derived from a thin
flux tube effective field theory~\cite{Luscher:1980ac}. We observe the Coulomb
couplings increase with decreasing sea quark mass for all three ans\"atze
considered herein. This trend, which is indicative of dynamical fermion
screening, has been previously observed for the standard potential
fits~\cite{Bali:2000vr}. It is interesting to see that this trend is replicated
in our modified fits as well, as it suggests that the modified Coulomb terms are
sensitive to the same short-distance physics as the standard ansatz.

\begin{table}[tb]
  \caption{\label{tab:SQP-alpha}The (effective) Coulomb term coefficients from the Cornell and modified fits to the untouched potentials.}
  \begin{ruledtabular}
    \begin{tabular}{ldddd}
$m_{\pi}\, (\si{MeV})$ & \mc{{\alpha}^{\rm cornell}_{\rm UT}} & \mc{\tilde{\alpha}^{\rm as}_{\rm UT}} & \mc{{\alpha}^{\rm sc}_{\rm UT}} \\
\midrule
Pure gauge & 0.286(7) & 0.293(7) & 0.301(7) \\
$701$ & 0.42(1) & 0.49(2) & 0.54(2) \\
$156$ & 0.61(4) & 0.72(6) & 0.84(7) 

    \end{tabular}
  \end{ruledtabular}
\end{table}

The crucial finding of this work is that the introduction of dynamical fermions
at any pion mass induces a measurable shift in the behaviour of centre vortices.
Applying the modified ans\"atze introduced herein, the pure gauge vortex-only
potential remains unable to reproduce the untouched string tension, whereas in
contrast the respective dynamical string tensions show good agreement.
The vortex-removed ensembles consistently show
complete removal of the long range confining potential. This reinforces the argument
that the salient non-perturbative properties of the ground state vacuum fields
are encapsulated in the centre vortex degrees of freedom.

%
%
%
%

\section{Conclusion}\label{sec:Conclusion}

In this work we have presented the first calculation of the static quark
potential from centre vortices obtained in the presence of dynamical fermions in
QCD. The difficulties in fitting a standard Coulomb term to a wide range of
vortex-removed values revealed a source of systematic contamination at moderate to
large separations, resulting in the under estimation of the untouched string
tension. In response we proposed two modified Coulomb ans\"atze. The first
modified ansatz seeks to model the effect of anti-screening in the running
coupling for QCD. The second modified ansatz takes the form of a Yukawa
potential, accomodating a dynamical effective gluon mass. Both ans\"atze for the
vortex-removed potential approach a constant value in the large $r$ limit, and
are able to describe the static quark potential on the
vortex-removed ensembles. Extending the modified Coloumb potentials with a
linear string tension enables fits to the untouched potential.

The vortex-removed ensembles lack a linear confining potential for both the
large and small pion masses considered here. Resolving the long-range behaviour
of the vortex-removed static quark potential with the fit parameter $\rho$
enables us to remove a source of systematic contamination in the untouched
potential fits, providing an improved determination of the untouched string
tension. In doing so, we find good agreement between the vortex-only and
untouched string tensions in the presence of dynamical fermions. The fact both
modified ans\"atze yield fit values for the string tension that are
essentially identical suggests that any systematic errors introduced by the
modifications are minimal. Evidence of quark loop screening is seen at the
light quark mass.

These results suggest that the presence of dynamical fermions resolves the
pure-gauge discrepancy between the original and vortex-only potential at large
distances, presenting an important step in understanding the QCD vacuum.
Historically, despite remarkable qualitative results, the centre-vortex model
has not agreed quantitatively with pure Yang-Mills calculations. It is
fascinating to see that with the improvements presented here that good agreement
is achieved for the string tension with the introduction of dynamical fermions
in full QCD. The mechanism for the observed phenomenological improvement is
currently unknown, and a direct examination of centre-vortex structure
complemented by probing of further quantities will assist in shedding light on
the complex relationship between centre vortices and the structure of the QCD
vacuum. Our findings strengthen the evidence that centre vortices are
responsible for the long-range confining potential of QCD, and provide a first
glimpse of the interplay between centre vortices and dynamical fermions.

Research to further explore centre vortices in full QCD is of interest, and will
be the subject of upcoming work. The relationship between dynamical fermions and
the geometry of centre vortices is also of interest, as it is well understood
that the confining potential of centre vortices arises from an area-law
percolating behaviour~\cite{Greensite:2016pfc,DelDebbio:1998luz,Dosch:1988ha}.
Use of a different operator basis in the variational analysis, particularly a
light meson operator, may also further clarify the long-range behaviour of the
vortex-modified potential and connections to string breaking.

\section{Acknowledgements}

We thank the PACS-CS Collaboration for making their 2 +1 flavour configurations
available via the International Lattice Data Grid (ILDG). This research was
undertaken with the assistance of resources from the National Computational
Infrastructure (NCI), provided through the National Computational Merit
Allocation Scheme and supported by the Australian Government through Grant No.
LE190100021 via the University of Adelaide Partner Share. This research is
supported by Australian Research Council through Grants No. DP190102215 and
DP210103706. WK is supported by the Pawsey Supercomputing Centre through the
Pawsey Centre for Extreme Scale Readiness (PaCER) program. W.K. would like to
thank Ross Young and Peter Tandy for valuable discussions. J.B. thanks Adam
Virgili for helpful discussions on the smearing of vortex-only configurations.

\bibliography{sqp_paper}

\end{document}